\documentclass[prd,preprint,tightenlines,showpacs,nofootinbib,superscriptaddress]{revtex4}
\usepackage{amsmath,latexsym,amssymb,amsthm,placeins,eufrak,color,amstext}

\DeclareMathAlphabet{\mathpzc}{OT1}{pzc}{m}{it}
\newcommand{\CurlyE}{\mathpzc{E}}

\newcommand{\ba}{\begin{eqnarray}}
\newcommand{\ea}{\end{eqnarray}}
\newcommand{\be}{\begin{equation}}
\newcommand{\ee}{\end{equation}}
\newcommand{\nn}{\nonumber}
\newcommand{\at}{\mathcal{T}}
\newcommand{\al}{\mathcal{L}}
\newcommand{\Et}{\mathcal{\cal E}}

\newcommand{\aA}{\mathcal{A}}
\newcommand{\ndd}{n}
\newcommand{\npi}{N}

\begin{document}

\title{From k-essence to generalised Galileons}

\author{C.~Deffayet} \email{deffayet@iap.fr}
\affiliation{AstroParticule \& Cosmologie,
UMR 7164-CNRS, Universit\'e Denis Diderot-Paris 7,
CEA, Observatoire de Paris,
10 rue Alice Domon et L\'eonie
Duquet, F-75205 Paris Cedex 13, France}

\author{Xian~Gao} \email{xgao@apc.univ-paris7.fr}
\affiliation{AstroParticule \& Cosmologie,
UMR 7164-CNRS, Universit\'e Denis Diderot-Paris 7,
CEA, Observatoire de Paris,
10 rue Alice Domon et L\'eonie
Duquet, F-75205 Paris Cedex 13, France}
\affiliation{Laboratoire de Physique Th\'{e}orique, \'{E}cole Normale Sup\'{e}rieure, 24 rue Lhomond, 75231 Paris, France}
\affiliation{Institut d'Astrophysique de Paris (IAP), UMR 7095-CNRS, Universit\'{e} Pierre et Marie Curie-Paris 6, 98bis Boulevard Arago, 75014 Paris, France}

\author{D.~A.~\surname{Steer}} \email{steer@apc.univ-paris7.fr}
\affiliation{AstroParticule \& Cosmologie,
UMR 7164-CNRS, Universit\'e Denis Diderot-Paris 7,
CEA, Observatoire de Paris,
10 rue Alice Domon et L\'eonie
Duquet, F-75205 Paris Cedex 13, France}

\author{G.~\surname{Zahariade}} \email{zahariad@apc.univ-paris7.fr}
\affiliation{AstroParticule \& Cosmologie,
UMR 7164-CNRS, Universit\'e Denis Diderot-Paris 7,
CEA, Observatoire de Paris,
10 rue Alice Domon et L\'eonie
Duquet, F-75205 Paris Cedex 13, France}

\begin{abstract}
{We determine the most general scalar field theories which have an action that depends on derivatives of order two or less, and have equations of motion that stay second order and lower on flat space-time. We show that those theories can all be obtained from linear combinations of Lagrangians made by multiplying a particular form of the Galileon Lagrangian by an arbitrary scalar function of the scalar field and its first derivatives.  We also obtain curved space-time extensions of those theories which have second order field equations for both the metric and the scalar field. This provide the most general extension, under the condition that field equations stay second order, of k-essence, Galileons, k-Mouflage as well as of the kinetically braided scalars. It also gives the most general action for a scalar classicalizer, which has second order field equations. We discuss the relation between our construction and the Euler hierachies of Fairlie {\it et al,} showing in particular that Euler hierachies allow one to obtain the most general theory when the latter is shift symmetric. As a simple application of our formalism, we give the covariantized version of the conformal Galileon.}
\end{abstract}

\date{\today}

\maketitle

\section{Introduction}

Scalar field models with derivative self interactions have attracted attention in various contexts. 
For instance, models of $k$-essence \cite{ArmendarizPicon:2000ah}  provide an interesting framework in which to investigate important early \cite{ArmendarizPicon:1999rj} as well as late time \cite{ArmendarizPicon:2000dh} issues of modern cosmology, while models similar to $k$-essence have been proposed in the context of relativistic MOND \cite{Bekenstein:1984tv}. All these models have the characteristic feature that their action depends solely on a scalar field $\pi$ and its first derivative --- clearly then, whatever the Lagrangian, the field equations stay second order. 

More recently, scalar models with actions depending on second derivatives of the fields have been considered, mainly inspired by the decoupling limit of the
Dvali-Gabadadze-Porrati (DGP) model and its cosmology \cite{Dvali:2000hr,COSMODGP} as well as the resulting modification of the gravitational interaction via the so-called Vainshtein mechanism \cite{Vainshtein:1972sx,Deffayet:2001uk}. Such models range from the  ``Galileon'' \cite{Nicolis:2008in}, to ``$k$-Mouflage" \cite{Babichev:2009ee} or ``kinetic gravity braided scalars" \cite{Deffayet:2010qz,Kobayashi:2010cm} and have different defining properties. A feature shared by the former and latter class of models, as well as some $k$-Mouflage models, is that they have an action which depends on second order derivatives of the fields. Hence, it is not {\it a priori} obvious how the field equations can stay second order, a property necessary in order to avoid propagating ghosts or extra degrees of freedom. This, however, can be achieved. The Galileon \cite{Nicolis:2008in}, for example, can be defined as the most general scalar theory which, in flat space-time, has field equations which are uniquely second order in derivatives. We note that in fact, Galileons were introduced rather earlier than \cite{Nicolis:2008in}, by Fairlie {\it et al.~} \cite{Fairlie} though in a different context. As we will also outline in section \ref{EULERHIER}, there the relevant Lagrangians were constructed through the successive application (called ``Euler hierachies") of the Euler-Lagrange operator to an arbitrary initial Lagrangian depending solely on the first derivative of a scalar field (with also the possibility of introducing arbitrary functions of the field first derivatives at intermediate steps). 

The curved space-time generalizations of those models are also interesting. 
As shown in Ref.~\cite{US}
the simplest covariantization of the original, four dimensional, Galileons led to field equations for the
scalar and its stress-tensor that contained third
derivatives. However, \cite{US} also showed how to eliminate
these higher derivatives by introducing suitable non-minimal,
curvature, couplings. Single scalar Galileons and their non minimal covariantization were further generalized to the multi-field (and $p$-forms) case, as well as to arbitrary dimensions, in Refs.~\cite{Deffayet:2009mn,Deffayet:2010zh,Padilla:2010de,Hinterbichler:2010xn}. More recently, \cite{deRham:2010eu,VanAcoleyen:2011mj} showed how to obtain the Galileons and covariant Galileons from  models with extra-dimensions.
Finally,  Ref.~\cite{Deffayet:2010qz} pointed out that a family of models which have Lagrangians depending linearly on  second derivatives of the fields, but also have second order field equations, have interesting properties when considered on curved space-time, due to an essential mixing between the scalar and the metric dubbed in \cite{Deffayet:2010qz}, kinetic gravity braiding.

The DGP model also generated new interest in massive gravity and its Vainshtein mechanism. This mechanism was first discussed \cite{Vainshtein:1972sx} in a simple non linear extension of the free theory for a massive graviton, the so-called Pauli-Fierz theory, as a way of getting rid of the bothering effects of the scalar polarization of the massive graviton, present for any non-vanishing graviton mass --- what is known as the van Dam-Veltman-Zakharov discontinuity \cite{vanDam:1970vg}. The Vainshtein mechanism, which was recently shown to work even in the simplest theories of massive gravity \cite{WORKVAIN}, can be attributed to the self-interactions of the scalar polarization of the graviton \cite{Deffayet:2001uk,ArkaniHamed:2002sp,Creminelli:2005qk,Deffayet:2005ys}. The latter interactions, which can be studied by taking an appropriate ``decoupling limit" \cite{ArkaniHamed:2002sp,Luty:2003vm}, take the form of derivative scalar self-couplings and the same is in fact true in the DGP model. 
For the Vainshtein mechanism to operate, however, there is no need to have field equations that are purely second order. This was shown explicitly in Ref.~\cite{Babichev:2009ee} in particular, which introduced a large family of scalar tensor models called $k$-Mouflage, and which used the Vainshtein mechanism to screen the effect of a scalar field at small distances. Note also that the Vainshtein mechanism also serves as one of the basis of the recently introduced ``classicalization" \cite{CLASSIC}. Finally, we stress that a recent attempt to obtain a massive gravity devoided of the unwanted Boulware Deser ghost \cite{Boulware:1973my} has a decoupling limit sharing crucial properties with some of the above mentioned scalar field models \cite{DRG}.

Hence, it is clear that scalar models which have derivative self interactions, possibly depending on second order derivatives, 
have numerous interesting properties. However, to our knowledge, these theories have so far not been extensively classified nor even constructed. It is the purpose of this work to do so. Namely, here, we will construct all theories of a scalar field $\pi$ in $D$ dimension and on flat space-time, which have actions depending on first and twice differentiated $\pi$'s as well as on undifferentiated $\pi$'s (hence without assuming necessarily a shift symmetry) but have field equations which stay of order two and lower. This will be carried out in section 
\ref{UNICSECTION} where our main result is first stated  and summarized (subsection \ref{SEC:RESULT}) before being proven.
We then show how to non minimally complete those theories in curved space-time, maintaining second order field equations of the scalar as well as for the metric (section \ref{COVAR}). Some examples are then discussed in relation with the Euler hierarchies construction, and we also illustrate our results giving the covariantization of the conformal Galileons (section \ref{EXAMPLES}). An introductory section \ref{GALIL} gathers some useful results.

\section{Galileon, kinetic gravity braiding and some useful results and notations}
\label{GALIL}

In this section we work in flat space only, and introduce the models studied in the remainder of this paper as well as some useful notation and results. We also revisit the Galileon model studied in \cite{Nicolis:2008in}.  Throughout we work in $D$ space-time dimensions, with signature $(-,+,+,\ldots)$.

\subsection{Introduction and two useful lemma}
\label{LEM1}

All models we consider depend on a single scalar field $\pi$ whose partial derivatives will be denoted by\footnote{Note that, when considered on curved space-time, $\pi_\mu$ will denote the covariant derivative acting on $\pi$, $\nabla_\mu \pi$, and so on for $\pi_{\mu \nu} ...$, i.e. partial derivatives are just to be replaced by covariant derivatives in the notation (\ref{defpimunu}).}
\be \label{defpimunu}
\pi_\mu \equiv \partial_\mu \pi \, , \qquad \pi_{\mu \nu} \equiv \partial_\mu \partial_\nu \pi \, , \qquad
\pi_{\mu \nu \alpha } \equiv \partial_\mu \partial_\nu \partial_\alpha \pi \, \qquad {\rm etc}.
\ee
Since derivatives commute on flat space-time, these tensors are symmetric under interchange of any indices.
The Lagrangians considered take the form
\ba \label{L0}
\al = \at_{(2n)}^{\mu_{\vphantom{()}1}
\ldots \mu_{\vphantom{()}n}\nu_{\vphantom{()}1}
\ldots \nu_{\vphantom{()}n} } \pi_{\mu_{\vphantom{()}1} \nu_{\vphantom{()}1}} \ldots \pi_{\mu_{\vphantom{()}n} \nu_{\vphantom{()}n}} \ ,
\ea
where $\at_{(2n)}$ is a $2n$-contravariant tensor function of $\pi$ and $\pi_\alpha$ only
\be
\at_{(2n)} = \at_{(2n)}(\pi,\pi_\alpha) \, .
\label{propT}
\ee
Note that the integer $n$ also denotes the number of twice differentiated $\pi$'s appearing in the Lagrangian (\ref{L0}).
Thus $\al = \al(\pi,\pi_\alpha,\pi_{\alpha \beta})$ and the corresponding field equations are ${\cal E}=0$, where 
\ba
 {\Et} &=& 
 \left[  \frac{\partial}{\partial \pi} - \partial_\mu \left(\frac{\partial}{\partial \pi_\mu} \right) +  \partial_\mu \partial_\nu \left(\frac{\partial }{\partial \pi_{\mu \nu}} \right) \right] {\cal L}
\label{EL}
\\
&\equiv&\hat{\Et}   {\cal L} \, .
\label{ELHAT}
\ea
 For future use, we begin by giving the {\it sufficient} conditions such that these equations are of order 2 or lower (in derivatives) on flat space-time. To do so, first note the following lemma:
\newline

\noindent {\bf Lemma. }{\it  The field equation derived from the Lagrangian $\mathcal{L}= {\cal T}_{(2)}^{\mu\nu}\pi_{\mu\nu}$
does not contain any derivative of order higher than 2. }\newline

This is straightforward to verify.  The second term in (\ref{EL}) yields one contribution in third derivatives of $\pi$, namely $\pi_{\alpha \beta \mu} \partial \at^{\alpha \beta}/\partial \pi_\mu$. An identical contribution arises from the last term in (\ref{EL}). However, given the relative sign in (\ref{EL}), these terms in third derivatives cancel.
This simple result is in fact at the basis of the model of kinetically braided scalar of Ref. \cite{Deffayet:2010qz}, where $\at^{\mu \nu}_{(2)} = f(\pi,\pi_\alpha) g^{\mu \nu}$ and $g_{\mu \nu}$ is the metric. In fact the above lemma generalizes to curved space-time as shown in appendix \ref{APPA} and used in section \ref{COVAR}. 

Now consider any $n > 1$. The second term in (\ref{EL}) again yields contributions to the equation of motion which are third order in derivatives. Those are cancelled by terms coming from the final term in Eq.~(\ref{EL}) by virtue of the above lemma. However, the final term in Eq.~(\ref{EL}) also yields ``dangerous terms'' (by which we mean terms of order three or more in derivatives) which are of the form $\pi_{\mu_{\vphantom{()}k} \mu_{\vphantom{()}l} \nu_{\vphantom{()}l}}$, $\pi_{\nu_{\vphantom{()}k} \mu_{\vphantom{()}l} \nu_{\vphantom{()}l}}$ or $\pi_{\nu_{\vphantom{()}k} \mu_{\vphantom{()}k} \mu_{\vphantom{()}l} \nu_{\vphantom{()}l}}$  where all the indices are contracted with those of $\at_{(2n)}$. Since derivatives commute on flat space-times, we immediately have the following result:
\newline

\noindent {\bf Main lemma.} {\it A sufficient condition for the field equations derived from the Lagrangian (\ref{L0}) to stay of order less or equal to 2 is that the tensor $\at_{(2n)}^{\mu_{\vphantom{()}1} \mu_{\vphantom{()}2}
\ldots \mu_{\vphantom{()}n}\nu_{\vphantom{()}1} \nu_{\vphantom{()}2}
\ldots \nu_{\vphantom{()}n}}$ is totally antisymmetric in its first $n$ indices $\mu_{\vphantom{()}i}$ as well as (separately) in its last $n$ indices $\nu_{\vphantom{()}i}$.}\newline

The main purpose of this paper is to study the converse of this simple result. Before doing so, and for future use, we first revisit the Galileon theory. As we will see, this provides a simple and fundamental example of the type of theory we will discuss.

\subsection{The flat space time Galileon revisited}

The starting point of Galileon models \cite{Nicolis:2008in,US} 
are Lagrangians 
${\cal L}^{\rm{Gal}}$ 
 of the form (\ref{L0}) with the tensor ${\cal T}_{(2n)}$ satisfying (\ref{propT}) as well as the properties of the main lemma above.   Furthermore, in flat space-time, they have equations of motion of order {\it strictly equal} to 2 (that is, they do not contain undifferentiated or once differentiated  $\pi$, but only twice differentiated $\pi$).

As we now outline, there exist several possible ways of writing the Galileon action:
each differs from the other by a different choice of tensor $\at_{(2n)}$ and a total derivative. However they lead to the same equations of motion and are hence equivalent.

To begin with, define the $2m$-contravariant tensor ${\mathcal A}_{(2m)}$ by
\begin{equation} \label{DEFAten}
\mathcal{A}_{(2m)}^{\mu_{\vphantom{()}1} \mu_{\vphantom{()}2}
\ldots \mu_{\vphantom{()}m} \nu_{\vphantom{()}1} \nu_{\vphantom{()}2}
\ldots \nu_{\vphantom{()}m}} \equiv
\frac{1}{(D-m)!}\,
\varepsilon^{\mu_{\vphantom{()}1}
\mu_{\vphantom{()}2}  \ldots
\mu_{\vphantom{()}m} \sigma_{\vphantom{()}1}\sigma_{\vphantom{()}2}\ldots
\sigma_{\vphantom{()}D-m}}_{\vphantom{\mu_{\vphantom{()}1}}}
\,\varepsilon^{\nu_{\vphantom{()}1} \nu_{\vphantom{()}2} \ldots
\nu_{\vphantom{()}m}}_{\hphantom{\nu_{\vphantom{()}1}
\nu_{\vphantom{()}2} \ldots
\nu_{\vphantom{()}2m}}\sigma_{\vphantom{()}1}
\sigma_{\vphantom{()}2}\ldots \sigma_{\vphantom{()}D-m}}
\end{equation}
where the totally antisymmetric Levi-Civita tensor is given by
\begin{equation} \label{DEFLC}
\varepsilon^{\mu_{\vphantom{()}1} \mu_{\vphantom{()}2} \ldots
\mu_{\vphantom{()}D}} = - \frac{1}{\sqrt{-g}}
\delta^{[\mu_{\vphantom{()}1}}_1 \delta^{\mu_{\vphantom{()}2}}_2
\ldots \delta^{\mu_{\vphantom{()}D}]}_D \, 
\end{equation}
with square brackets denoting unnormalized permutations.  (The definitions (\ref{DEFAten}) and (\ref{DEFLC}) are also valid in arbitrary curved space-times with metric $g_{\mu \nu}$ and $D\geq m$.)
Thus $\aA_{(2m)}$ is antisymmetric
in its first $m$ indices as well as, separately, in its last $m$ indices.  Other useful properties of $\aA_{(2m)}$ are given in Appendix \ref{PROPA}.

A first possible Lagrangian for the Galileon
is given by \cite{Deffayet:2009mn}
\ba \label{LGAL1}
{\cal L}^{\rm{Gal},1}_{\npi} &=&\left( {\mathcal A}_{(2\ndd+2)}^{\mu_{\vphantom{()}1}
\ldots \mu_{\vphantom{()}\ndd+1}\nu_{\vphantom{()}1}
\ldots \nu_{\vphantom{()}\ndd+1} } \pi_{\mu_{n+1}} \pi_{\nu_{n+1}} \right) \pi_{\mu_{\vphantom{()}1} \nu_{\vphantom{()}1}} \ldots \pi_{\mu_{\vphantom{()}\ndd} \nu_{\vphantom{()}\ndd}}
\nonumber
 \\
&\equiv & \at^{\mu_{\vphantom{()}1}
\ldots \mu_{\vphantom{()}\ndd}  \nu_{\vphantom{()}1}
\ldots \nu_{\vphantom{()}\ndd} }_{(2\ndd),\rm{Gal},1}  \pi_{\mu_{\vphantom{()}1} \nu_{\vphantom{()}1}} \ldots \pi_{\mu_{\vphantom{()}\ndd} \nu_{\vphantom{()}\ndd}},
\ea
with
\ba
\at^{\mu_{\vphantom{()}1}\label{TENSGAL1}
\ldots \mu_{\vphantom{()}n}  \nu_{\vphantom{()}1}
\ldots \nu_{\vphantom{()}n} }_{(2\ndd),\rm{Gal},1}  &\equiv &  {\mathcal A}_{(2\ndd+2)}^{\mu_{\vphantom{()}1}
\ldots \mu_{\vphantom{()}\ndd+1}\nu_{\vphantom{()}1}
\ldots \nu_{\vphantom{()}\ndd+1} } \pi_{\mu_{n+1}} \pi_{\nu_{n+1}} \, .
\ea
Here and henceforth $\npi$ indicates the number of $\pi$'s appearing in the Lagrangian of a given Galileon model
so that
\be 
\npi= \ndd+2.
\nonumber
\ee
As discussed in \cite{Deffayet:2009mn}, the Lagrangian ${\cal L}^{\rm{Gal},1}_{\npi}$ also reads:
\ba \label{nico1}
{\cal L}^{\rm{Gal},1}_{\npi}&=& - \sum_{\sigma \in S_{\ndd+1}} \epsilon(\sigma)
\bigl[ \pi_{\vphantom{\mu_1}}^{\mu_{\sigma(1)}} \pi_{\mu_1}\bigr]
\bigl[\pi_{\hphantom{\mu_{\sigma(2)}} \mu_2}^{\mu_{\sigma(2)}}
\pi_{\hphantom{\mu_{\sigma(3)}} \mu_3}^{\mu_{\sigma(3)}} \ldots
\pi_{\hphantom{\mu_{\sigma(\ndd+1)}} \mu_{\ndd+1}}^{\mu_{\sigma(\ndd+1)}}\bigr], \nonumber \\&=& \label{nico1bis}
-\sum_{\sigma \in S_{\ndd+1}} \epsilon(\sigma)
g^{\mu_{\sigma(1)}\nu_{\vphantom{()}1}}
g^{\mu_{\sigma(2)}\nu_{\vphantom{()}2}} \ldots
g^{\mu_{\sigma(\ndd+1)}\nu_{\vphantom{()}\ndd+1}}
(\pi_{\nu_1} \pi_{\mu_1})
(\pi_{\nu_2 \mu_2} \pi_{\nu_3 \mu_3}\ldots \pi_{\nu_{\ndd+1} \mu_{\ndd+1}}),
\ea
where $\sigma$ denotes a permutation of signature $\epsilon(\sigma)$ of the permutation group $S_{\ndd+1}$, and in order for the Lagrangian to be non-vanishing,
\be
\ndd+1 \leq D \qquad \Longleftrightarrow \qquad N \leq D+1\,.
\label{MAXX}
\ee 
This is the original form presented in \cite{Nicolis:2008in}, and the equality of (\ref{LGAL1}) and  (\ref{nico1bis}) can be seen \cite{Deffayet:2009mn} using the identity 
\begin{equation}
\sum_{\sigma \in S_D} \epsilon(\sigma)
g^{\mu_{\sigma(1)}\nu_{\vphantom{()}1}}
g^{\mu_{\sigma(2)}\nu_{\vphantom{()}2}} \ldots
g^{\mu_{\sigma(D)}\nu_{\vphantom{()}D}} =
- \varepsilon^{\mu_{\vphantom{()}1} \mu_{\vphantom{()}2}
\ldots \mu_{\vphantom{()}D}}\,
\varepsilon^{\nu_{\vphantom{()}1} \nu_{\vphantom{()}2}
\ldots \nu_{\vphantom{()}D}}.
\end{equation}
Using (\ref{EL}), the field equations derived from the Lagrangian ${\cal L}^{{\rm Gal},1}_{\npi}$ read 
\be
\Et = -\npi \times {\cal E}_{\npi}=0 \, ,
\ee
where 
\ba
{\cal E}_{\npi} &=&-\sum_{\sigma \in S_{\ndd+1}} \epsilon(\sigma)
\prod_{i=1}^{\ndd+1} \pi_{\hphantom{\mu_{\sigma(i)}}
\mu_i}^{\mu_{\sigma(i)}}, \nonumber
\\
&=& {\mathcal A}_{(2\ndd+2)}^{\mu_{\vphantom{()}1}
\ldots \mu_{\vphantom{()}\ndd+1}\nu_{\vphantom{()}1}
\ldots \nu_{\vphantom{()}\ndd+1} }
\pi_{\mu_1\nu_1} \pi_{\mu_2\nu_2} \ldots \pi_{\nu_{\ndd+1} \mu_{\ndd+1}}.
\label{eofmG}
\ea
These are only second order, as advertised.  Notice that the index $\npi$ on $\Et_\npi$ indicates that it is the equation of motion coming from ${\cal L}^{\rm{Gal},1}_{\npi}$ (which contains $\npi$ factors of $\pi$): thus $\Et_\npi$ contains $N-1$ factors of $\pi$. 
The Galileon model with the largest number of fields in $D$ dimensions has $N=D+1$.  In this case, ${\cal E}_{D+1}$ is simply proportional to the determinant of the Hessian, the matrix of second derivatives $\pi_{\mu \nu}$. As such, the equation ${\cal E}_{D+1} = 0 $ is known  as the Monge-Amp\`ere equation, and it has various interesting properties, in particular in relation to integrability (see e.g.~\cite{Fairlie:1994in}).  At the same time 
\be
{\cal L}^{{\rm Gal},1}_{D+1} \propto \det
\left(\begin{array}{cc}\pi_{\mu \nu} & \pi_{\nu} \\ \pi_{\mu} & 0\end{array}\right) \, ,
\nn
\ee
which is the left-hand-side of the Bateman equation \cite{Fairlie:1994in,Bateman}.

 Finally, 
in $D$ dimensions, the total Galileon Lagrangian is given by a linear combination of Lagrangians ${\cal L}_{N}^{{\rm Gal},1}$ with $N=2, \cdots, D+1$.
In $D=4$ dimensions, these are simply the 4 terms given in \cite{Nicolis:2008in}.

A second possible Lagrangian for the Galileon with, again, $\npi = n+2$ fields is given by
\ba \label{LGAL2}
{\cal L}^{\rm{Gal},2}_{\npi} &=&\left( {\mathcal A}_{(2\ndd)}^{\mu_{\vphantom{()}1}
\ldots \mu_{\vphantom{()}\ndd}\nu_{\vphantom{()}1}
\ldots \nu_{\vphantom{()}\ndd} } \pi_{\mu_1} \pi_{\lambda} \pi^{\lambda}_{\hphantom{\lambda} \nu_1} \right) \pi_{\mu_{\vphantom{()}2} \nu_{\vphantom{()}2}} \ldots \pi_{\mu_{\vphantom{()}\ndd} \nu_{\vphantom{()}\ndd}},\\
&\equiv & \at^{\mu_{\vphantom{()}1}
\ldots \mu_{\vphantom{()}\ndd}  \nu_{\vphantom{()}1}
\ldots \nu_{\vphantom{()}\ndd} }_{(2\ndd),\rm{Gal},2} \pi_{\mu_{\vphantom{()}1} \nu_{\vphantom{()}1}} \ldots \pi_{\mu_{\vphantom{()}\ndd} \nu_{\vphantom{()}\ndd}},
\label{EXEQ21}
\ea
where (see also section \ref{ANTI})
\ba
\at^{\mu_{\vphantom{()}1}
\ldots \mu_{\vphantom{()}\ndd}  \nu_{\vphantom{()}1}
\ldots \nu_{\vphantom{()}\ndd} }_{(2\ndd),\rm{Gal},2} = \frac{1}{n}
\,
 {\mathcal A}_{(2\ndd)}^{\alpha_{\vphantom{()}1}
\ldots \alpha_{\vphantom{()}\ndd}\nu_{\vphantom{()}1}
\ldots \nu_{\vphantom{()}\ndd} }
 && \Big[\left( \pi^{\mu_1} \pi_{\alpha_1}\right) \delta^{\mu_2}_{\; \; \alpha_2} \ldots \delta^{\mu_n}_{\; \; \alpha_n}  \nonumber \\
&& + \delta^{\mu_1}_{\; \;  \alpha_1} \left( \pi^{\mu_2} \pi_{\alpha_2} \right) \delta^{\mu_3}_{\; \;  \alpha_3} \ldots \delta^{\mu_n}_{\; \;  \alpha_n} \nonumber \\
&& + \ldots \nonumber \\
&& + \delta^{\mu_1}_{\; \;  \alpha_1} \ldots \delta^{\mu_{n-1}}_{\; \;  \alpha_{n-1}} \left( \pi^{\mu_n} \pi_{\alpha_n} \right) \Big].
\label{TT2}
\label{TGAL2}
\ea

Finally, the third form of interest is given by
\ba \label{LGAL3}
{\cal L}^{{\rm Gal},3}_{\npi} &= & \left( {\mathcal A}_{(2\ndd)}^{\mu_{\vphantom{()}1}
\ldots \mu_{\vphantom{()}\ndd}\nu_{\vphantom{()}1}
\ldots \nu_{\vphantom{()}\ndd} } \pi_{\lambda} \pi^{\lambda}\right) \pi_{\mu_{\vphantom{()}1} \nu_{\vphantom{()}1}} \ldots \pi_{\mu_{\vphantom{()}\ndd} \nu_{\vphantom{()}\ndd}}
\ea
so that
\ba
\at^{\mu_{\vphantom{()}1}
\ldots \mu_{\vphantom{()}\ndd}  \nu_{\vphantom{()}1}
\ldots \nu_{\vphantom{()}\ndd} }_{(2\ndd),{\rm Gal},3}  &=& X \mathcal{A}_{(2\ndd)}^{\mu_{\vphantom{()}1}
\ldots \mu_{\vphantom{()}\ndd}   \nu_{\vphantom{()}1}
\ldots \nu_{\vphantom{()}\ndd}  }
\ea
where\footnote{Note that in the context of $k$-inflation and other models, $X$ is often defined to be $-\pi_\mu \pi^\mu/2$. In order to simplify our equations, the factor of $-1/2$ is not included here.}
\be
X\equiv \pi_\mu \pi^\mu \, .
\label{Xdef}
\ee

The three Lagrangians (\ref{LGAL1}), (\ref{LGAL2}) and (\ref{LGAL3}) are in fact all equal up to a total derivative. Indeed, on defining $J^\mu_{\npi}$ by
\ba \label{DEFJ}
J^\mu_{\npi}= X \mathcal{A}_{(2n)}^{\mu \mu_{2}\cdots\mu_{n}\nu_{1}\nu_{2}\cdots\nu_{n}}\pi_{\nu_{1}}\pi_{\mu_{2}\nu_{2}}\cdots\pi_{\mu_{n}\nu_{n}} \, ,
\ea
it follows directly that
\ba
\mathcal{L}_{\npi}^{\rm{Gal},2}&=&-\frac{1}{2}\mathcal{L}_{\npi}^{\rm{Gal},3}+\frac{1}{2} \partial_\mu J^\mu_{\npi} \, .
\label{REL2}
\ea
Furthermore on using the properties of $\aA_{(2n)}$ given in appendix \ref{PROPA}, it follows  that
\ba{\label{3Lid}}
(\npi-2)\mathcal{L}_{\npi}^{\rm{Gal},2}=\mathcal{L}_{\npi}^{\rm{Gal},3}-\mathcal{L}_{\npi}^{\rm{Gal},1} \, .
\ea
Thus we also have
\ba
\mathcal{L}_{\npi}^{\rm{Gal},1}&=&\frac{\npi}{2}\mathcal{L}_{\npi}^{\rm{Gal},3}-\frac{\npi-2}{2} \partial_\mu J^\mu_{\npi}, \label{REL1}\\
\mathcal{L}_{\npi}^{\rm{Gal},1}&=&-\npi \mathcal{L}_{\npi}^{\rm{Gal},2}+\partial_\mu J^\mu_{\npi} \, . 
\label{REL3}
\ea
From (\ref{REL2}), (\ref{REL1}) and (\ref{REL3}) it therefore follows that the equations of motion of all three Galileon Lagrangians are identical, given by (\ref{eofmG}), and strictly  of second order. 

Finally, observe from (\ref{eofmG}) and (\ref{LGAL3}) that ${\cal L}^{\rm{Gal},3}_{\npi}$ can be rewritten as
\be
{\cal L}^{\rm{Gal},3}_{\npi} = X {\cal E}_{\npi-1} 
\label{DEFGAL3}
\ee
where ${\cal E}_{\npi-1}$ are the equations of motion coming from ${\cal L}^{\rm{Gal}}_{\npi-1}$ (where we drop the index $1,2,3$). In this form, it is manifest that Galileon models containing a given number $\npi$ of $\pi$ fields can be obtained from the field equations of the same models with one less field. This property, though implicit in \cite{Nicolis:2008in}, is very well explained by the hierachical construction of \cite{Fairlie} which preceeded by far Ref.~\cite{Nicolis:2008in} and discussed first, as far as we know, what are called here and elsewhere Galileons (see \cite{Fairlie:2011md}).  This  hierachical construction will be discussed in section \ref{EXAMPLES}.

\subsection{Galileon Lagrangians in terms of cycles}

The three equivalent Galileon Lagrangians presented above all satisfy the {sufficient} conditions of the main lemma of section \ref{LEM1}.  In order to study the {necessary} conditions it will be useful to introduce a new notation consisting of the cycles $[i]$ and $\langle i \rangle$.

We define $[i]$ by
\ba \label{DEFBIS[}
[i] &\equiv & \pi^{\mu_1}_{\hphantom{\mu_1} \mu_2} \pi^{\mu_2}_{\hphantom{\mu_2}{\mu_3}} \pi^{\mu_3}_{\hphantom{\mu_3}\mu_4} \cdots \pi^{\mu_{i}}_{\hphantom{\mu_{i}}\mu_1},
\ea
so that for example
\be
[1]=\Box \pi \, , \qquad [2] = \pi^{\alpha}_{\hphantom{\alpha} \beta} \pi^{\beta}_{\hphantom{\beta}\alpha}  \, .
\ee
Similarly,
\ba \label{DEFBIS<}
\left<i\right> &\equiv & \pi_{\mu_1} \pi^{\mu_1}_{\hphantom{\mu_1} \mu_2} \pi^{\mu_2}_{\hphantom{\mu_2}{\mu_3}} \pi^{\mu_3}_{\hphantom{\mu_3}\mu_4} \cdots  \pi^{\mu_{i}}_{\hphantom{\mu_{i}}\mu_{i+1}}\pi^{\mu_{i+1}},
\ea
so that
\be
\left<1 \right> =  \pi_{\alpha} \pi^{\alpha}_{\hphantom{\alpha} \beta} \pi^{\beta}
\ee
Note that $[i]$ contains $i$ factors of $\pi$ as well as $i$ twice-differentiated $\pi$'s, whereas $\left<i\right>$ contains $i+2$ factors of $\pi$, but again $i$ twice-differentiated $\pi$'s.

Using this notation, and in the case of $\npi = 4$ fields, the three Galileon Lagrangians can be written as 
\begin{eqnarray}
{\cal L}_{N=4}^{{\rm Gal},1} &=& - \left(\Box \pi\right)^2 \left(\pi_{\mu}\,\pi^{\mu}\right)
+ 2 \left(\Box \pi\right)\left(\pi_{\mu}\,\pi^{\mu\nu}\,\pi_{\nu}\right) \, 
\nonumber 
\\
&& + \left(\pi_{\mu\nu}\,\pi^{\mu\nu}\right) \left(\pi_{\rho}\,\pi^{\rho}\right)
-2 \left(\pi_{\mu}\pi^{\mu\nu}\,\pi_{\nu\rho}\,\pi^{\rho}\right) \, 
\nonumber 
\\
&=& X \left(  [2]  - [1]^2 \right) + 2 \left( [1]\left<1 \right> -  \left<2 \right> \right) \, ,
\nonumber 
\\
{\cal L}_{N=4}^{{\rm Gal},2} &=& - \left(\Box \pi\right) \pi_\mu \pi^{\mu \nu} \pi_\nu +  \pi_\mu \pi^{\mu \nu} \pi_{\nu \rho} \pi^\rho  =    \left<2\right> - [1] \left<1\right>   \, ,
\nonumber 
 \\
{\cal L}_{N=4}^{{\rm Gal},3} &=&  \left(\pi_\lambda \pi^\lambda\right) \left(  \pi_{\mu \nu} \pi^{\mu \nu} - \left(\Box \pi\right)^2 \right)
=  X \left(  \left[2\right] - \left[1\right]^2\right) \, .
\nonumber 
\ea

Furthermore it will be useful to define
\ba
\left[
\begin{array}{cccc}
p_1 &  p_2 & \cdots & p_r  \\1 & 2 & \cdots & r
\end{array}\right] &=& [1]^{p_1} [2]^{p_2} \cdots [r]^{p_r} \label{DEF[},
\ea
as well as 
\ba
\left<
\begin{array}{cccc}
q_1 &  q_2 & \cdots & q_s  \\1 & 2 & \cdots & s
\end{array}\right> &=& \left<1\right>^{q_1} \left<2\right>^{q_2} \cdots \left<s\right>^{q_s}\label{DEF<},
\ea
where the $p_i$ and $q_i$ are positive (or vanishing) integers.\footnote{Note that the right hand sides of equations (\ref{DEF[}) and (\ref{DEF<}) are uniquely specified respectively by the ordered sets $(p_1,p_2,\cdots,p_r)$ as well as $(q_1,q_2,\cdots,q_s)$. Hence our notation on the left hand side of equations (\ref{DEF[}) and (\ref{DEF<}) is a bit redundant, but we feel this will ease the reading of some of the equations below.}  Using this notation it follows from (\ref{TT2}) that ${\cal L}_{N}^{{\rm Gal},2}$  can be expressed as a sum of terms $\left[
\begin{array}{cccc}
p_1&  p_2 & \cdots & p_r  \\1 & 2 & \cdots & r
\end{array}\right]\left<
\begin{array}{ccc}
q_1 &   \cdots & q_s  \\1 &  \cdots&s
\end{array}\right>$
with $\sum_{j=1}^{s} {q_j}=1$. (See above in case $N=4$.)

Thus, for example, the equations of motion coming ${\cal L}_{N=4}$ are 
\begin{eqnarray}
 0 = {\cal E}_{N=4}&=& 
  -  \left(\Box \pi \right)^3
-2 \left(\pi_{\mu}^{\hphantom{\mu}\nu}\,\pi_{\nu}^{\hphantom{\nu}\rho}\,\pi_{\rho}^{\hphantom{\rho}\mu}\right)
+3 \left(\Box \pi\right) \left(\pi_{\mu\nu}\, \pi^{\mu\nu} \right) 
\nonumber 
\\
&=& -  [1]^3 - 2 [3] + 3 [1][2] 
\nonumber 
\\
&=& - \left[\begin{array}{c}3 \\1 \end{array} \right] - 2 \left[\begin{array}{c}1 \\3 \end{array} \right] + 3 \left[\begin{array}{cc}1 &1\\1&2 \end{array} \right] \, .
\label{E4a}
\end{eqnarray}
More generally, for any $N$, ${\cal E}_{\npi}$ can be expressed as a specific linear combination of monomials  of the form (\ref{DEF[}):
\ba \label{COEFC}
{\cal E}_{\npi} = \sum {\cal C}^{\npi}_{p_1,\cdots,p_r} \left[
\begin{array}{cccc}
p_1 &  p_2 & \cdots & p_r  \\1 & 2 & \cdots & r
\end{array}\right],
\ea
where the sum runs over the $r-$uplets $(p_1,\cdots,p_r)$ verifying the constraint
\ba
\npi= 1+\sum_{i=1}^{i=r} i \times p_i ,
\ea
and
${\cal C}^{\npi}_{p_1,\cdots,p_r}$ are real coefficients which do not depend on the dimension of space-time, but only depend on the number of field $\npi$, explicitly
\begin{eqnarray*}
\mathcal{C}_{p_{1},p_{2},\cdots p_{r}}^{N} 
& = & 
\left(-1\right)^{N+p_{1}+p_{2}+\cdots + p_{r}}\frac{\left(N-1\right)!}{\left(p_{1}!p_{2}!\cdots p_{r}!\right)1^{p_{1}}2^{p_{2}}\cdots r^{p_{r}}}.
\end{eqnarray*}

\section{Uniqueness theorem in flat space-time}
\label{UNICSECTION}\label{UNIK}
\subsection{The result}
\label{SEC:RESULT}
We are now in the position to study the converse of the main lemma of section \ref{LEM1}. However, before presenting the proof in section \ref{PROOF}, we first state our result.
Namely, in flat space-time, the most general theory satisfying the three conditions
\begin{itemize}
\item[(i)] its Lagrangian contains derivatives of order 2 or less of the scalar field $\pi$;
\item[(ii)] its Lagrangian is polynomial in the second derivatives of $\pi$;
\item[(iii)] the corresponding field equations are of order 2 or lower in derivatives
\end{itemize}
has a Lagrangian which is given by an arbitrary linear combination of the Lagrangians ${\cal L}_{\ndd}\{f\}$ (each containing $\ndd$ of twice differentiated $\pi$) of the form
\ba
{\cal L}_{\ndd}\{f\}&=& f(\pi,X) \times {\cal L}^{\rm{Gal},3}_{\npi = \ndd + 2}, \nonumber \\
&=& f(\pi,X) \times \left( X {\mathcal A}_{(2 \ndd)}^{\mu_{\vphantom{()}1}
\ldots \mu_{\vphantom{()}\ndd}\nu_{\vphantom{()}1}
\ldots \nu_{\vphantom{()}\ndd} }  \pi_{\mu_{\vphantom{()}1} \nu_{\vphantom{()}1}} \ldots \pi_{\mu_{\vphantom{()}\ndd} \nu_{\vphantom{()}\ndd}}\right). \label{GENL}
\ea
Here $f(\pi,X)$ is an arbitrary scalar function of $\pi$ and $X$, generally different for each $n$, and the braces in ${\cal L}_{\ndd}\{f\}$ denote that ${\cal L}_\ndd$ is a functional of $f$. The equations of motion corresponding to each  ${\cal L}_{\ndd}\{f\}$ are\footnote{We use the notation $f_{X} \equiv f_{,X}$, $f_{\pi} \equiv f_{,\pi}$ and so on.}
\begin{eqnarray}
0 & = & 2\left(f+Xf_{X}\right)\mathcal{E}_{N}+4\left(2f_{X}+Xf_{XX}\right)\mathcal{L}_{N+1}^{\text{Gal},2}\nonumber \\
 &  & +X\left[2Xf_{X\pi}-\left(n-1\right)f_{\pi}\right]\mathcal{E}_{N-1} \nonumber \\
 && -n\left(4Xf_{X\pi}+4f_{\pi}\right)\mathcal{L}_{N}^{\text{Gal},2} - nXf_{\pi\pi} \mathcal{L}_{N-1}^{{\rm Gal},1}.\label{EOFM}\end{eqnarray}
where $N=n+2$.  Notice the dependence of these equations on $\pi^{\mu \nu}$ as well as $\pi^\mu$ (when $f$ is non-constant).

\subsection{Proof of uniqueness} 
\label{PROOF}

In flat space-time, the only scalar quantities which are polynomial in second 
derivatives 
of $\pi$ must be constructed from $[i]$ and $\left<i\right>$, defined respectively  in Eqs.~(\ref{DEFBIS[}) and (\ref{DEFBIS<}). Recall that these both contain $i$ times a twice differentiated $\pi$.
Hence, the most general scalar theory obeying conditions (i) and (ii) has a Lagrangian which is a linear combination of monomials, each of the form ${\cal L}^{p_1,p_2,\cdots, p_r}_{q_1,q_2,\cdots,q_s}$  defined by
\ba \label{MONO}
{\cal L}^{p_1,p_2,\cdots, p_r}_{q_1,q_2,\cdots,q_s} = f(\pi,X) \times \left[
\begin{array}{cccc}
p_1 &  p_2 & \cdots & p_r  \\1 & 2 & \cdots & r
\end{array}\right]\left<
\begin{array}{cccc}
q_1 &  q_2 & \cdots & q_s  \\1 & 2 & \cdots & s
\end{array}\right>
\ea
where $f$ is an arbitrary scalar function of $\pi$ and $X$ (different for each monomial ${\cal L}^{p_1,p_2,\cdots, p_r}_{q_1,q_2,\cdots,q_s}$), and the other quantities have been defined in (\ref{DEF[}) and (\ref{DEF<}).  Formally then
\be
{\cal L} = \sum_{\{p_i\}\{q_j\}} C_{\{p_i\}\{q_j\}} {\cal L}^{p_1,p_2,\cdots, p_r}_{q_1,q_2,\cdots,q_s}
\label{SUMC}
\ee
where the sum is over the set of $\{p_i\} =(p_1,\cdots, p_r)$ and $\{q_j\} =(q_1,\cdots, q_s)$, and (the ratio of) the constant coefficients $C_{\{p_i\}\{q_i\}}$ will be determined below.
Note also that the number $\npi$ of fields, and the number $\ndd$ of twice differentiated $\pi$ which appear in a given product
$\left[
\begin{array}{cccc}
p_1 &  p_2 & \cdots & p_r  \\1 & 2 & \cdots & r
\end{array}\right]\left<
\begin{array}{cccc}
q_1 &  q_2 & \cdots & q_s  \\1 & 2 & \cdots & s
\end{array}\right> $ are given respectively by
\ba
\npi &=& \left(\sum_{i=1}^{i=r} p_i \times i\right)  +\left(\sum_{j=1}^{j=s} q_j \times (j+2)\right),
\label{NPI} \\
\ndd &=& \npi- 2 \sum_{j=1}^{j=s}q_j.
\label{NDD}
\ea
We now look for the most general theory which obeys condition (iii) as well as (i) and (ii).

\subsubsection{Fourth order derivatives containing $\Box \pi^{\alpha}{}_{\beta}$}

Inspired by the proof given in \cite{Nicolis:2008in}, start from a particular monomial of the form
${\cal L}^{p_1,p_2,\cdots, p_r}_{q_1,q_2,\cdots,q_s}$, given in (\ref{MONO}), with specific values of the $p_i$  (for $1\leq i\leq r$) and of the $q_j$ (for $1\leq j\leq s$). If the theory considered obeys condition (iii), {\it all} third and fourth order derivatives must disappear from the field equations. There are many 4th order derivative terms, but first focus on those containing $\Box \pi^{\alpha}{}_{\beta}$ (others will be discussed later).  When varying ${\cal L}^{p_1,p_2,\cdots, p_r}_{q_1,q_2,\cdots,q_s}$, such terms can appear through
\ba
\delta_{\pi} \; [i]^{p_i} &\supset&  \frac{2 i p_{i}}{i-1} [\Box(i-1)] [i]^{p_i - 1} \; \delta \pi \qquad \qquad (i>1)
\label{WAY1}
\\
\delta_{\pi} \; \langle i \rangle^{q_i} &\supset& 2 q_i  \langle \Box(i-1) \rangle  \langle i \rangle^{q_i - 1}\; \delta \pi\qquad \qquad (i>1)
\label{WAY2}
\ea
where 
\ba
[\Box(j)] &\equiv  & \sum_{k=1}^j \pi^{\mu_1}_{\hphantom{\mu_1}{\mu_2}} \pi^{\mu_2}_{\hphantom{\mu_2}\mu_3} \cdots  \pi^{\mu_{k-1}}_{\hphantom{\mu_{k-1}}\mu_k} (\Box \pi^{\mu_{k}}_{\hphantom{\mu_{k}}\mu_{k+1}}) \cdots
  \pi^{\mu_{j-1}}_{\hphantom{\mu_{j-1}}\mu_j} \pi^{\mu_j}_{\hphantom{\mu_j} \mu_1} 
\\
&=& j\times \pi^{\mu_1}_{\hphantom{\mu_1}{\mu_2}} \pi^{\mu_2}_{\hphantom{\mu_2}\mu_3} \cdots  \pi^{\mu_{j-1}}_{\hphantom{\mu_{j-1}}\mu_j} (\Box \pi^{\mu_{j}}_{\hphantom{\mu_{j}}\mu_{1}}) \\
\langle \Box(j) \rangle & \equiv & 
\sum_{k=1}^j   \pi_{\mu_1}  \pi^{\mu_1}_{\hphantom{\mu_1}{\mu_2}} \cdots  \pi^{\mu_{k-1}}_{\hphantom{\mu_{k-1}}\mu_k} (\Box \pi^{\mu_{k}}_{\hphantom{\mu_{k}}\mu_{k+1}}) \cdots
  \pi^{\mu_{j}}_{\hphantom{\mu_{j}}\mu_{j+1}} \pi^{\mu_{j+1}} \, .
\ea
Now consider the contribution from (\ref{WAY1}). On varying ${\cal L}^{p_1,p_2,\cdots, p_r}_{q_1,q_2,\cdots,q_s}$ this yields the term
\ba \label{4DER1}
\frac{2 i p_i}{i-1} f(\pi,X) \times [1]^{p_1} [2]^{p_2} \cdots 
[i-1]^{p_{i-1}} 
[\Box(i-1)] [i]^{p_i-1}  \cdots [r]^{p_r}\left<
\begin{array}{cccc}
q_1 &  q_2 & \cdots & q_s  \\1 & 2 & \cdots & s
\end{array}\right>
\ea
 in the equations of motion, and it can
only be cancelled if one adds to the Lagrangian a term proportional to
\ba
f \times \left[
\begin{array}{ccccccccc}
p_1+1 &  p_2 & \cdots & p_{i-2}& p_{i-1}+1&p_{i}-1&p_{i+1}&\cdots& p_r  \\1 &  2 & \cdots & i-2& i-1 &i&i+1&\cdots& r
\end{array}\right]\left<
\begin{array}{cccc}
q_1 &  q_2 & \cdots & q_s  \\1 & 2 & \cdots & s
\end{array}\right>.
\label{NOTHING}
\ea
Indeed, variation of the first term $[1]^{p_1+1} \equiv \left(\Box \pi \right)^{p_1+1}$ gives, after integrating by parts and shifting the $\Box$ operator onto one of the cycles $[i-1]$, a term proportional to  (\ref{4DER1}). The same term is obtained from varying one of the twice differentiated $\pi$ inside a cycle $[i-1]$ and, on integrating by parts, acting with the derivatives on one $\Box \pi$.
Thus a {\it necessary} condition for the theory considered to obey conditions  (i), (ii) and (iii) is that it must contain in its action the specific linear combination 
\ba
& f& \left<
\begin{array}{cccc} 
q_1 &  q_2 & \cdots & q_s  \\1 & 2 & \cdots & s
\end{array}\right> \times
\nn
\\
&&\; \;  \left\{  \left[
\begin{array}{cccc}
p_1 &  p_2 & \cdots & p_{r}\\1 &  2 & \cdots & r
\end{array}\right]
+
\alpha_{[\; ]}
\left[
\begin{array}{ccccccccc}
p_1+1 &  p_2 & \cdots & p_{i-2}& p_{i-1}+1&p_{i}-1&p_{i+1}&\cdots& p_r  \\1 &  2 & \cdots & i-2& i-1 &i&i+1&\cdots& r
\end{array}\right]\right\}
\label{TERM2}
\ea
where $\alpha_{[ \; ]}$, which is nothing other than the ratio of two of the $C$ coefficients defined in (\ref{SUMC}),
is given by
\be
\alpha_{[\; ]} =  - \frac{ i p_i}{(p_1+1)(p_{i-1} + 1)(i-1)} \, .
\nn
\ee
Notice that $\alpha_{[ \; ]}$ is independent of $f(\pi,X)$.
It is straightforward to check that the two terms in factor of $f$ in (\ref{TERM2}) each have the same number $\npi$ of fields, and number $\ndd$ of twice differentiated $\pi$, as given in (\ref{NPI}) and (\ref{NDD}).

A similar reasoning can be applied to the ``$\left<\right>$'' piece in (\ref{MONO}). Indeed, using (\ref{WAY2}) it follows that any theory obeying conditions (i), (ii) and (iii), and which has a term in its action given by
${\cal L}^{p_1,p_2,\cdots, p_r}_{q_1,q_2,\cdots,q_s}$, must also contain a term 
\ba \label{TERM3}
\alpha_{\langle \rangle} \times f \times \left[
\begin{array}{cccc}
p_1+1 &  p_2 & \cdots & p_r  \\1 & 2 & \cdots & r
\end{array}\right]\left<
\begin{array}{cccccccc}
q_1 &   \cdots & q_{j-1}+1&q_j-1&q_{j+1}&\cdots & q_s  \\1 &  \cdots & j-1&j&j+1&\cdots&s
\end{array}\right>
\ea
where
\be
\alpha_{\langle \rangle} = - \frac{ q_j}{(p_1+1)(q_{j-1} + 1)} \, .
\nn
\ee
 Once again observe that the above terms (\ref{MONO}) and (\ref{TERM3}) have the same number $\npi$ of fields, and same number $\ndd$ of twice differentiated $\pi$'s.

We can thus define mappings $F$ and $G$ on the set of monomials (\ref{MONO})
which appear in the Lagrangian of any theory obeying conditions (i), (ii) and (iii), such that $F$ maps any term
${\cal L}^{p_1,p_2,\cdots, p_r}_{q_1,q_2,\cdots,q_s}$, i.e.~(\ref{MONO}), to the monomial (\ref{4DER1}) and similarly $G$ maps the term (\ref{MONO}) to (\ref{TERM3}). Then, any two terms related by those mapping (or their inverse $F^{-1}$ and $G^{-1}$) must have coefficients which are equal (up to combinatorial factors) in order to eliminate fourth order derivatives containing $\Box \pi^{\alpha}{}_{\beta}$. These mapping can easily be pictured by a graph whose nodes are labelled by the set $\{p_1,p_2,\cdots,p_r,q_1,q_2,\cdots,q_s\}$ and represent the monomial, and such that two nodes are connected if and only if they are image of each other by the mappings $F$ or $G$ or their inverse (or successive applications of $F$,$G$, $F^{-1}$ or $G^{-1}$). 

Now observe that in going from (\ref{MONO}) to (\ref{NOTHING}), the power of the cycle $[i]$ is lowered by 1, whereas the power of the cycles $[i-1]$ and $[1]$ is increased by 1. Similarly, in going from the term (\ref{MONO}) to (\ref{TERM3}) the power of the cycle $\langle j \rangle $ is lowered by 1 whereas that of the cycles $\langle j-1 \rangle $ and $[1]$ increases by 1. Hence, by acting recursively with the mappings $F$ and $G$, starting from the cycles $[r]$ and $\left<s\right>$  which have the {\it largest} length $r$ and $s$ respectively, one ends up with the conclusion that {\it any term ${\cal L}^{p_1,p_2,\cdots, p_r}_{q_1,q_2,\cdots,q_s}$ is connected (after several applications of the maps $F$ and $G$) to a term which has all but $p_1$ and $q_1$ vanishing}. That is, a term of the form
\ba 
{\cal L}^p_q
&=& f \times [1]^p \left<1\right>^q
\nn
\\
&=& f \times \left(\Box \pi \right)^p \left(\pi^\mu \pi_{\mu \nu} \pi^\nu\right)^q \, ,
 \label{TERM5}
\ea
where
\ba
q &=& \sum_{j=1}^{s} q_j \, ,
\label{Q}
\\
p &=& \sum_{i=1}^{r} (i p_i) + \sum_{j=1}^{s} q_j (j-1) \, .
\label{P}
\ea
Alternatively, on using (\ref{NPI}) and (\ref{NDD}), $p$ and $q$ are determined by the number of fields and twice differentiated $\pi$ in the part of the monomial containing second derivatives, through
\ba
p &=& \frac{1}{2} \left( 3 \ndd-\npi\right), \\
q&=& \frac{1}{2}\left(\npi - \ndd\right).
\label{Q2}
\ea
Since the term (\ref{TERM5}) is the {\it same} for all ${\cal L}^{p_1,p_2,\cdots, p_r}_{q_1,q_2,\cdots,q_s}$ which has fixed values of $\npi$  and $\ndd$, one can conclude that the graph introduced above, and representing monomials with fixed values of $\npi$ and $\ndd$, is {\it connected}. In other words, it means that any term ${\cal L}^{p_1,p_2,\cdots, p_r}_{q_1,q_2,\cdots,q_s}$ with fixed $\npi$  and $\ndd$ must appear in the action, with a common function $f$ and fixed coefficients.

Hence, to conclude the first part of our reasoning,  we get a family of (possibly trivial but each uniquely determined) theories indexed by the values of $\npi$ and $\ndd$. More specifically, for a given values of $\npi$ and $\ndd$ the theory is uniquely specified by the coefficient in front of (\ref{TERM5}) (as well as the function $f$). The coefficients of all the other monomials with the same number of $\npi$ and $\ndd$ will be proportional to the coefficient of (\ref{TERM5}) and the proportionality factor will be independent of the specific function $f$ (since this factor is uniquely determined by  requiring the vanishing of terms with four derivatives $\Box \pi^{\alpha}{}_{\beta}$, and this procedure is blind to the chosen form for $f$). 

\subsubsection{Other fourth order derivative terms}

Now consider other fourth order derivative terms which may appear in the equations of motion. In particular, if $q \geq 2$, the field equation derived from (\ref{TERM5}) will contain a fourth derivative term proportional to
\ba
f \times \left(\Box \pi \right)^p \left(\pi^\lambda \pi^\rho \pi^\sigma \pi^\tau \pi_{\lambda \rho \sigma \tau} \right)\left(\pi^\mu \pi_{\mu \nu} \pi^\nu\right)^{q-2}
\ea
(obtained by varying one of the twice differentiated $\pi$ appearing in one cycle $\left<1\right>$ and integrating by parts on the other cycle $\left<1\right>$). However, it is impossible to cancel this term by the variation of any of the other terms in the Lagrangian (connected to ${\cal L}^{p}_{q}$ through $F$ and $G$). Thus we conclude that there are two possibilities --- the power $q$ appearing in (\ref{TERM5}) must take the value $0$ or $1$, so that on using (\ref{Q2}),
\ba
q=0 &\Longleftrightarrow& \npi =\ndd
\label{POS1}
\\
q=1 & \Longleftrightarrow& \npi =\ndd + 2.
\label{POS2}
\ea

\subsubsection{Second order equations of motion}

First focus on $q=0$, namely one particular theory which obeys (i), (ii), (iii), has a fixed value of $\ndd$, and $\npi=\ndd$.  From (\ref{Q}), it follows that all the $q_j$ must vanish so that the Lagrangian is a sum of monomials made of $\left[
\begin{array}{cccc}
p_1&  p_2 & \cdots & p_r  \\1 & 2 & \cdots & r
\end{array}\right]$ only.   As we have just seen, these have relative coefficients which are fixed and independent of the choice of the function $f$. One can conclude that those coefficients must be the ones, ${\cal C}^{\npi+1}_{p_1,\cdots,p_r}$, appearing in the expansion of ${\cal E}_{\npi+1}$ (see Eq.~(\ref{COEFC})).
Indeed, we know that if we consider ${\cal E}_{\npi+1}$ as an action, this action has a vanishing equation of motion since ${\cal E}_{\npi+1}$ is a total derivative. Hence it obeys the hypotheses (i), (ii) and (iii) and it has also the correct power, $\ndd = \npi$,  of twice differentiated $\pi$ appearing in the expansion in terms of monomials $\left[
\begin{array}{cccc}
p_1&  p_2 & \cdots & p_r  \\1 & 2 & \cdots & r
\end{array}\right]$.
Hence, one is led to the conclusion that any theory obeying (i), (ii) and (iii), with a fixed value of $\ndd$ and $\npi= \ndd$ must have a Lagrangian of the form $g \times \mathcal{E}_{\npi+1}$, where $g$ is some function of $\pi$ and $X$. Thus, on using (\ref{DEFGAL3}), it can be rewritten as (defining $f\equiv g X^{-1}$)
\ba 
\label{FAM1}
\mathcal{L}_{\ndd}^{(3)}\{f\}& \equiv &f(\pi,X) \times \mathcal{L}^{{\rm Gal},3}_{\ndd+2}
\\
&=& f(\pi,X) \times \mathcal{L}^{{\rm Gal},3}_{\npi} \, .
\nn
\ea

A similar argument applies to the other family of models. Here $q=1=\sum_{j=1}^{s} {q_j}$, and $\npi=\ndd+2$ for a given value of $\ndd$: these models must have a Lagrangian of the form
\ba \label{FAM2}
\mathcal{L}_{\ndd}^{(2)}\{f\}&\equiv &f(\pi,X) \times \mathcal{L}^{{\rm Gal},2}_{\ndd+2}, \\
&=&f(\pi,X) \times \mathcal{L}^{{\rm Gal},2}_{\npi}.
\nn
\ea
Indeed $\mathcal{L}^{{\rm Gal},2}_{\npi}$ obeys hypotheses (i), (ii) and (iii), and, as recalled after equation (\ref{DEF<}), has an expansion in terms of monomials $\left[
\begin{array}{cccc}
p_1&  p_2 & \cdots & p_r  \\1 & 2 & \cdots & r
\end{array}\right]\left<
\begin{array}{ccc}
q_1 &   \cdots & q_s  \\1 &  \cdots&s
\end{array}\right>$
with $\sum_{j=1}^{j=s} {q_j}=1$.

Finally, it is straightforward to show that the field equations derived from Lagrangians (\ref{FAM1}) and (\ref{FAM2}) are indeed second order (so that third order derivative terms also vanish).
This is in fact a direct consequence of the ``main lemma'' of the previous section, since using the expressions (\ref{LGAL2}) and (\ref{LGAL3}) one can see that Lagrangians (\ref{FAM1}) and (\ref{FAM2}) are of the form (\ref{L0}).  In the case of ${\cal L}^{(3)}_{n}$, the explicit equations of motion are given in (\ref{EOFM}).

\subsubsection{Unique family of models satisfying the ``main lemma''}

Finally, we are in a position to show that (\ref{FAM1}) and (\ref{FAM2}) are in fact equivalent up to total derivatives, so that the unique theory satisfying (i), (ii) and (iii) is indeed given by an arbitrary linear combination of the Lagrangians ${\cal L}^{(3)}_{\ndd}\{f\}$
 as advertised in section \ref{SEC:RESULT}.
 
Define $\mathcal{L}_{n}^{(1)}\{f\} \equiv f(\pi,X) {\cal{L}}_{n+2}^{\rm{Gal}, 1}$. Then the identity (\ref{3Lid}) amongst the 3 Galileon models is now generalised to
\be
{\label{3Lidgen}}
        n\,\mathcal{L}_{n}^{(2)}\{f\}=\mathcal{L}_{n}^{(3)}\{f\}-\mathcal{L}_{n}^{(1)}\{f\} \, ,
\ee
and similarly (\ref{REL2}) becomes 
    \be
    {\label{3Libp}}
        2{\cal L}_{n}^{(2)}\left\{ f+Xf_{X}\right\} =-{\cal L}_{n-1}^{(1)}\left\{ Xf_{\pi}\right\} -{\cal L}_{n}^{(3)}\left\{ f\right\} +\partial_{\mu}\left(f(\pi,X)J_{(n)}^{\mu}\right),
    \ee 
$J^\mu_{(n)}\equiv J^\mu_N$ is given in Eq.~(\ref{DEFJ}). Elimination of $\mathcal{L}^{(1)}_{n}$ between these two equations yields a recurrence relation between $\mathcal{L}^{(2)}_{n}$ and $\mathcal{L}^{(3)}_{n}$, namely
    \be
    {\label{L2frec}}
        {\cal L}_{n}^{(2)}\left\{ f\right\} =-\left(n-1\right){\cal L}_{n-1}^{(2)}\left\{ \frac{\partial g_{1}}{\partial\pi}\right\} +{\cal L}_{n}^{(3)}\left\{ \frac{g_{1}}{X}\right\} +{\cal L}_{n-1}^{(3)}\left\{ \frac{\partial g_{1}}{\partial\pi}\right\} + \text{tot.~div.}\;,
    \ee
with
    \begin{equation}
    \nn
        g_{1}\{f\} =-\frac{1}{2}\int_{0}^{X}dYf\left(\pi,Y\right).
    \end{equation}
On using (\ref{L2frec}) repeatedly, it follows that (up to a total derivative) ${\cal L}_{n}^{(2)}\{f\}$ can be expressed
as linear combination of ${\cal L}_{i}^{(3)}\{f\}$.  Specifically 
    \ba{\label{L2rec}}
    {\cal L}_{n}^{(2)}\{f\}={\cal L}_{0}^{(3)}\left\{\frac{\partial g_{n,1}}{\partial\pi}\right\}+
    \sum_{i=1}^{n-1} {\cal L}_{i}^{(3)}\left\{\frac{g_{n,i}}{X}+\frac{\partial g_{n,i+1}}{\partial\pi}\right\}+{\cal L}_{n}^{(3)}\left\{\frac{g_{n,n}}{X}\right\}+\text{tot.~div.}\;,
    \ea
where $\mathcal{L}_0^{(3)}\{f\}=Xf$ for consistency, and
    \begin{eqnarray}
        g_{n,i}\{f\} &\equiv& \frac{\left(n-1\right)!}{\left(i-1\right)!}g_{n-i+1}\{f\},
        \nn
        \\
        {g}_{i}\{f\} &\equiv& - \frac{1}{2^{i}}\left(\frac{\partial}{\partial\pi}\right)^{i-1}
        \int_{X_0}^{X}dX_{1}\int_{X_0}^{X_{1}}dX_{2}\cdots\int_{X_0}^{X_{i-1}}dX_{i}\,f\left(\pi,X_{i}\right),
        \nn
    \end{eqnarray}
 Thus Eq.~(\ref{L2rec}) shows the equivalence of the $q=0$ and $q=1$ Lagrangians given in (\ref{FAM1}) and (\ref{FAM2}) respectively.

Finally, observe that $\mathcal{L}_{D}^{(3)}\{f\}$ is in fact a linear combination of $\mathcal{L}^{(3)}_k$ with $k=0,\ldots,D-1$. This follows from (\ref{L2rec}) together with the fact that for $n=D$ equation (\ref{3Lidgen}) reduces to $D\,\mathcal{L}_{D}^{(2)}\{f\}=\mathcal{L}_{D}^{(3)}\{f\}$.  Thus to conclude, the most general Lagrangian in $D$-dimensions obeying conditions (i), (ii) and (iii) is given by
\be
{\cal L} = \sum_{n=0}^{D-1} {\cal L}_{n}\{ f_n \} \, ,
\label{final}
\ee
where $f_n$ are arbitrary functions of $\pi$ and $X$ and 
${\cal L}_{\ndd}\{f\} \equiv  {\cal L}_{\ndd}^{(3)}\{ f \}$.

\section{Covariantization} 
\label{COVAR}

We now turn our discussion from flat space-time to curved space-time.
Let us start from the general Lagrangian ${\cal L}_{n}\{f\}$ in our family (\ref{GENL}), writing it in the form of Eq.~(\ref{L0}) with ${\cal T}_{(2n)}$ defined by
\ba
\mathcal{T}_{(2n)}&=& \mathcal{T}_{(2n)}\left(\pi,X\right)
\nn
\\
&=&f(\pi,X)\times X \mathcal{A}_{(2n)} ,
\ea
and where $n$ is the number of second derivatives of the field in the Lagrangian. On replacing all partial derivatives appearing in this Lagrangian by covariant derivatives, we obtain a minimally covariantized theory. As we will now see, following \cite{Deffayet:2009mn}, the field equations of this covariantized model contain derivatives of order higher than two (varying the action with respect to $\pi$ or with respect to the metric): however, we will also see that there exists a non minimal covariantization removing all such higher order derivatives.

Let us then consider variation with respect to the scalar $\pi$ of the minimally covariantized version of (\ref{GENL}).
It reads 
\be
\delta\mathcal{L}_{n}\{f\} =\delta \mathcal{T}_{(2n)}^{\mu_{1}\cdots\mu_{n}\nu_{1}\cdots\nu_{n}}\pi_{\mu_{1}\nu_{1}}\cdots\pi_{\mu_{n}\nu_{n}}+n\mathcal{T}_{(2n)}^{\mu_{1}\cdots\mu_{n}\nu_{1}\cdots\nu_{n}}\delta\pi_{\mu_{1}\nu_{1}}\pi_{\mu_{2}\nu_{2}}\cdots\pi_{\mu_{n}\nu_{n}}.
\label{DELC}
\ee
Above, the only ``dangerous terms" (recall that these are terms leading to expressions in the field equations depending on derivatives of order higher than two) come only from the second term on the right hand side. Indeed by virtue of a straightforward generalisation to curved space-time of the first lemma in section \ref{LEM1} (see the proof in appendix \ref{APPA}), dangerous terms coming from the first piece on the right hand side of (\ref{DELC}) are exactly compensated by terms coming from the second piece. After these compensations, first derivatives of the Riemann tensor (i.e.~third derivatives of the metric) remain, 
and are therefore troublesome. Indeed we have
\be
\delta\mathcal{L}_{n}\{f\}
\sim -\frac{n(n-1)}{4}\mathcal{T}_{(2n)}^{\mu_{1}\cdots\mu_{n}\nu_{1}\cdots\nu_{n}}\pi^{\lambda}R_{\mu_{1}\mu_{2}\nu_{1}\nu_{2};\lambda}\pi_{\mu_{3}\nu_{3}}\cdots\pi_{\mu_{n}\nu_{n}}\delta\pi,
\ee
where here and below we use the same notation as in \cite{Deffayet:2009mn} so that
the symbol $\sim$ means that we write only the dangerous terms, up to total derivatives.

In order for the equations of motion to be second order, it is enough to add a finite number of terms whose variations exactly cancel the third derivatives of the metric.  Starting with the above dangerous term, we  can add a term  proportional to $\left(\int_{X_{0}}^{X}\mathcal{T}_{(2n)}^{\mu_{1}\cdots\mu_{n}\nu_{1}\cdots\nu_{n}} \left(\pi,X_1\right) dX_{1}\right)R_{\mu_{1}\mu_{2}\nu_{1}\nu_{2}}\pi_{\mu_{3}\nu_{3}}\cdots\pi_{\mu_{n}\nu_{n}}$ to our Lagrangian.  Then we need to add another term in order to compensate for the dangerous terms arising from our correction, and so on. The general  term in this series of Lagrangians is
 \be
\mathcal{L}_{n,p}\{ f\}=\mathcal{P}_{(p)}^{\mu_{1}\mu_{2}\cdots\mu_{n}\nu_{1}\nu_{2}\cdots\nu_{n}}\mathcal{R}_{(p)}\mathcal{S}_{(q\equiv n-2p)}
\ee
where we again follow the notation of \cite{Deffayet:2009mn} and 
\ba
\mathcal{R}_{(p)}&\equiv &\prod_{i=1}^{p}R_{\mu_{2i-1}\mu_{2i}\nu_{2i-1}\nu_{2i}}, 
\nn
\\
 \mathcal{S}_{(q\equiv n-2p)}&\equiv&\prod_{i=0}^{q-1}\pi_{\mu_{n-i}\nu_{n-i}},
 \nn
 \ea 
 and $\mathcal{P}_{(p)}$ is the ($p$ times) repeated integral of $\mathcal{T}_{(2n)}$ with respect to $X$ defined by\footnote{$X_0$ being an arbitrary constant, whose presence is related to the possibility 
 of adding terms, all vanishing in flat space, that avoid higher derivatives. See e.g. (\ref{L4CGALC})-(\ref{L5CGALC}).}
\ba
\mathcal{P}_{(p)} \equiv \int_{X_{0}}^{X} dX_1 \int_{X_{0}}^{X_{1}} dX_2 \cdots\int_{X_{0}}^{X_{p-1}} dX_p  \; \mathcal{T}_{(2n)}^{\mu_{1}\mu_{2}\cdots\mu_{n}\nu_{1}\nu_{2}\cdots\nu_{n}}\left(\pi,X_1\right).
\nn
\ea
Notice that we use the conventions $\mathcal{L}_{n,0}\{f\}=\mathcal{L}_{n}\{f\}$, ${\cal{S}}_{1}=\pi_{\mu_n \nu_n} $ and ${\cal{S}}_{q\leq 0} = 1$. It follows from the lemma of Appendix \ref{APPA} that $\mathcal{L}_{n,p}\{f\}$ does not yield any more dangerous terms for $p \geq \lfloor n/2 \rfloor$ (where $\lfloor n/2 \rfloor$ denotes the integer part of $n/2$). Thus, as we will confirm, only a finite number of terms  $\mathcal{L}_{n,p}\{f\}$ is necessary.
 In order to make sure that adding these terms to the initial Lagrangian exactly cancels all higher derivatives in the equations of motion, and to determine the coefficients that make this possible, we now compute their  variation $\delta\mathcal{L}_{n,p}$ with respect to the field $\pi$, only paying attention to dangerous terms.  We get (after suitable integrations by parts)
\ba
\delta\mathcal{L}_{n,p} &=&\delta \mathcal{P}_{(p)}^{\mu_{1}\cdots\mu_{n}\nu_{1}\cdots\nu_{n}}\mathcal{R}_{(p)}\mathcal{S}_{(q)}+(n-2p)\mathcal{P}_{(p)}^{\mu_{1}\cdots\mu_{n}\nu_{1}\cdots\nu_{n}}\mathcal{R}_{(p)}\mathcal{S}_{(q-1)}\delta\pi_{\mu_{2p+1}\nu_{2p+1}}\nonumber \\
&=&2\mathcal{P}_{(p-1)}^{\mu_{1}\cdots\mu_{n}\nu_{1}\cdots\nu_{n}}\pi^{\lambda}\delta\pi_{\lambda}\mathcal{R}_{(p)}\mathcal{S}_{(q)}+(n-2p)\mathcal{P}_{(p)}^{\mu_{1}\cdots\mu_{n}\nu_{1}\cdots\nu_{n}}\mathcal{R}_{(p)}\mathcal{S}_{(q-1)}\delta\pi_{\mu_{2p+1}\nu_{2p+1}}\nonumber \\
&\sim& -2p\mathcal{P}_{(p-1)}^{\mu_{1}\cdots\mu_{n}\nu_{1}\cdots\nu_{n}}\mathcal{R}_{(p-1)}\mathcal{S}_{(q)}\pi^{\lambda}R_{\mu_{2p-1}\mu_{2p}\nu_{2p-1}\nu_{2p};\lambda}\delta\pi\nonumber \\
&& -\frac{(n-2p)(n-2p-1)}{4}\mathcal{P}_{(p)}^{\mu_{1}\cdots\mu_{n}\nu_{1}\cdots\nu_{n}}\mathcal{R}_{(p)}\mathcal{S}_{(q-2)}\pi^{\lambda}R_{\mu_{2p+1}\mu_{2p+2}\nu_{2p+1}\nu_{2p+2};\lambda}\delta\pi. \nonumber
\ea
Notice that, once again, the third derivatives of $\pi$ coming from the variation of $\mathcal{P}_{(p)}$ are exactly canceled by those coming from the variation of $\mathcal{S}_{(q)}$. The second Bianchi identity {\it ensures} that no other derivatives of the Riemann tensor appear. To summarise, in order that the equations of motion contain no more than second order derivatives, the Lagrangian must be given by the linear combination
\be
\mathcal{L}^{\rm {cov}}_{n}\{f\}=\sum_{p=0}^{\lfloor\frac{n}{2}\rfloor}\mathcal{C}_{n,p}\mathcal{L}_{n,p}\{f\},
\label{THEANSWER}
\ee
where the coefficients obey the specific recurrence relation $\mathcal{C}_{n,p+1}=-\frac{(n-2p)(n-2p-1)}{8(p+1)}\mathcal{C}_{n,p}$ for $p\geq 0$ and $\mathcal{C}_{n,0}=1$. 
This gives
\ba \label{COEFCNP} \mathcal{C}_{n,p}=\left(-\frac{1}{8}\right)^{p}\frac{n!}{(n-2p)!p!}.\ea
Note that similar expressions have been obtained in \cite{Deffayet:2009mn}. However, there, the covariantization was derived for the Lagrangians ${\cal L}^{Gal,1}_N$. It was also observed that the covariantized action could be written in different forms using various total derivatives. The same is true here and hence, taking also into account (\ref{REL1}), it is not straightforward to compare the covariantized form (\ref{THEANSWER})-(\ref{COEFCNP}) to the one given in \cite{Deffayet:2009mn}.

 Remarkably  the Lagrangian $\mathcal{L}^{{\rm cov}}_{n}\{f\}$ also yields second order equations for the metric. Indeed by computing the variation of each term appearing in the above linear combination, but this time with respect to the  metric $g_{\mu \nu}$ (denoting the metric variation by $\delta g_{\mu \nu}$), we obtain (given that the tensor $\mathcal{P}_{(p)}$ does not depend on derivatives of the metric)
\be
\delta\mathcal{L}_{n,p} \sim p\mathcal{P}_{(p)}^{\mu_{1}\cdots\mu_{n}\nu_{1}\cdots\nu_{n}}\mathcal{R}_{(p-1)}\delta R_{\mu_{2p-1}\mu_{2p}\nu_{2p-1}\nu_{2p}}\mathcal{S}_{(q)}+(n-2p)\mathcal{P}_{(p)}^{\mu_{1}\cdots\mu_{n}\nu_{1}\cdots\nu_{n}}\mathcal{R}_{(p)}\mathcal{S}_{(q-1)}\delta\pi_{\mu_{2p+1}\nu_{2p+1}}
\nonumber \ee
 with (see equations (31) and (32) in Ref. \cite{Deffayet:2009mn})
\ba
\delta\pi_{\mu_{2p+1}\nu_{2p+1}}&=&-\frac{1}{2}\pi^{\lambda}(\delta g_{\lambda\mu_{2p+1};\nu_{2p+1}}+\delta g_{\lambda\nu_{2p+1};\mu_{2p+1}}-\delta g_{\mu_{2p+1}\nu_{2p+1};\lambda})
\nonumber 
\\
\mathcal{P}_{(p)}^{\mu_{1}\cdots\mu_{n}\nu_{1}\cdots\nu_{n}}\delta R_{\mu_{2p-1}\mu_{2p}\nu_{2p-1}\nu_{2p}}&=&2\mathcal{P}_{(p)}^{\mu_{1}\cdots\mu_{n}\nu_{1}\cdots\nu_{n}}\delta g_{\mu_{2p-1}\nu_{2p};\mu_{2p}\nu_{2p-1}}
\nn
\\ && \qquad \qquad \qquad +\mathcal{P}_{(p)}^{\mu_{1}\cdots\mu_{n}\nu_{1}\cdots\nu_{n}}\delta g^{\sigma}{}_{\mu_{2p-1}}R_{\sigma\mu_{2p}\nu_{2p-1}\nu_{2p}} \nonumber.
\ea
This yields 
\ba
\delta\mathcal{L}_{n,p} &\sim& 2p\mathcal{P}_{(p)}^{\mu_{1}\cdots\mu_{n}\nu_{1}\cdots\nu_{n}}\mathcal{R}_{(p-1)}\mathcal{S}_{(q)}\delta g_{\mu_{2p-1}\nu_{2p};\mu_{2p}\nu_{2p-1}} 
\nonumber \\ && 
+\frac{(n-2p)}{2}\mathcal{P}_{(p)}^{\mu_{1}\cdots\mu_{n}\nu_{1}\cdots\nu_{n}}\mathcal{R}_{(p)}\mathcal{S}_{(q-1)}\pi^{\lambda}\delta g_{\mu_{2p+1}\nu_{2p+1};\lambda}.  \nonumber 
\ea
After two integrations by parts, we are led to
\ba
\delta\mathcal{L}_{n,p}&\sim& 4p\mathcal{P}_{(p-1)}^{\mu_{1}\cdots\mu_{n}\nu_{1}\cdots\nu_{n}}\mathcal{R}_{(p-1)}\mathcal{S}_{(q)}\pi^{\lambda}\pi_{\nu_{2p-1}\mu_{2p}\lambda}\delta g_{\mu_{2p-1}\nu_{2p}}\nonumber \\
&& -\frac{p(n-2p)}{2}\mathcal{P}_{(p)}^{\mu_{1}\cdots\mu_{n}\nu_{1}\cdots\nu_{n}}\mathcal{R}_{(p-1)}\mathcal{S}_{(q-1)}\pi^{\lambda}R_{\mu_{2p}\mu_{2p+1}\nu_{2p-1}\nu_{2p+1};\lambda}\delta g_{\mu_{2p-1}\nu_{2p}} \nonumber \\
&& -\frac{p(n-2p)}{2}\mathcal{P}_{(p)}^{\mu_{1}\cdots\mu_{n}\nu_{1}\cdots\nu_{n}}\mathcal{R}_{(p-1)}\mathcal{S}_{(q-1)}\pi^{\lambda}R_{\mu_{2p-1}\mu_{2p}\nu_{2p-1}\nu_{2p};\lambda}\delta g_{\mu_{2p+1}\nu_{2p+1}}\nonumber \\
&& -\frac{(n-2p)(n-2p-1)}{2}\mathcal{P}_{(p)}^{\mu_{1}\cdots\mu_{n}\nu_{1}\cdots\nu_{n}}\mathcal{R}_{(p)}\mathcal{S}_{(q-2)}\pi^{\lambda}\pi_{\lambda\mu_{2p+2}\nu_{2p+2}}\delta g_{\mu_{2p+1}\nu_{2p+1}} \, .
\nonumber 
\ea
On relabeling $\mu_{2p-1}\leftrightarrow\mu_{2p+1}$ and $\nu_{2p}\leftrightarrow\nu_{2p+1}$ in the second term,  the second and third terms are seen to cancel. Also, by changing $\nu_{2p}\leftrightarrow\nu_{2p-1}$ in the first term, it is clear that all the other dangerous terms vanish if the coefficients $\mathcal{C}_{n,p}$ are given by Eq.~(\ref{COEFCNP}). 

To conclude, the generalisation to curved space-time of our family of Lagrangians (\ref{GENL}) which yield second order equations of motion, is given by (\ref{THEANSWER}).

\section{Examples}
\label{EXAMPLES}
Although in section \ref{UNIK} we have completely classified all models of the form  (\ref{L0}) (and more generally all models obeying conditions (i), (ii) and (iii) of section \ref{UNIK}), it is of interest to discuss some specific models in which the tensor $\at_{(2n)}$ of Eqs.(\ref{L0}-\ref{propT}) takes a simple form. We will in turn give a straightforward method with which to construct a suitable tensor $\at_{(2n)}$ --- namely satisfying the properties given in the the main lemma of section \ref{LEM1}.  We then discuss the hierachical constructions of \cite{Fairlie} (see also \cite{Fairlie:2011md} for a recent summary) in our notation. Finally, we give the covariantization of the conformal Galileon in 4 dimensions (section \ref{CONFGALCOV}).

\subsection{Construction with antisymmetric tensors}

One way to obtain a tensor $\at_{(2n)}$ is to consider a set of twice contravariant tensors $F^{\mu \nu}_{i}$, $i = 1 \ldots n$, (and $n \leq D$) depending on $\pi$ and $\pi^\lambda$  (where $\mu$ and $\nu$ label space-time indices, whereas $i$ labels the tensors). Then to define the  $D$-contravariant tensor $\CurlyE_{\{F_i\}}$ by 
\ba 
\label{DEFE}
\CurlyE_{\{F_i\}}^{\hphantom{\{F_i\}}\mu_{\vphantom{()}1} \ldots \mu_{\vphantom{()}n} \sigma_{\vphantom{()}1} \ldots \sigma_{\vphantom{()}D-n}} \equiv
\,\varepsilon_{\nu_{\vphantom{()}1}  \ldots
\nu_{\vphantom{()}n}}^{\hphantom{nu_{\vphantom{()}1}  \ldots
\nu_{\vphantom{()}n}}
\sigma_{\vphantom{()}1} \ldots
\sigma_{\vphantom{()}D-n}} F_{(1}^{\mu_{1}\nu_{1}} F_{2}^{\mu_2 \nu_2} \ldots F_{n)}^{\mu_n\nu_n}
\ea
where the brackets denote symmetrisation over the tensors $F^{\mu \nu}_{i}$. The tensor $\CurlyE_{\{F_i\}}$ is easily seen to be completely antisymmetric over its first $n$ indices as well as (separately) on it last $D-n$ indices. 

The special choice
\ba \label{MET}
F^{\mu \nu}_i = g^{\mu \nu}, \qquad {\rm for \; all} \; i
\ea
gives a tensor $\CurlyE$ equal to the Levi-Civita $\varepsilon$ tensor. Another interesting case arises by choosing, for example, $F_1^{\mu \nu}=\pi^\mu \pi^\nu$, and all the other $F^{\mu \nu}_i$ given by the metric as in (\ref{MET}). In that case the tensor $\CurlyE$ will be denoted by $\tilde{\varepsilon}_{(n)}$
and is given by
\ba
\tilde{\varepsilon}_{(n)}{}^{\mu_{\vphantom{()}1} \ldots \mu_{\vphantom{()}n} \sigma_{\vphantom{()}1} \ldots \sigma_{\vphantom{()}D-n}} = \frac{1}{n}
\,\varepsilon_{\nu_{\vphantom{()}1}  \ldots
\nu_{\vphantom{()}n}}^{\hphantom{nu_{\vphantom{()}1}  \ldots
\nu_{\vphantom{()}n}}
\sigma_{\vphantom{()}1} \ldots
\sigma_{\vphantom{()}D-n}} && \left(\pi^{\mu_1} \pi^{\nu_1} g^{\mu_2 \nu_2} \ldots g^{\mu_n \nu_n} \right. \nonumber \\
&& + g^{\mu_1 \nu_1} \pi^{\mu_2} \pi^{\nu_2} g^{\mu_3 \nu_3} \ldots g^{\mu_n \nu_n} \nonumber \\
&& + \ldots \nonumber \\
&& + \left.g^{\mu_1 \nu_1} \ldots g^{\mu_{n-1} \nu_{n-1}} \pi^{\mu_n} \pi^{\nu_n} \right).
\ea
This tensor enters in particular in the form (\ref{LGAL2})-(\ref{TT2}) of the action for the Galileon, as we will see below. We can also construct a tensor $\CurlyE$ given by a linear combination of $\varepsilon$ and $\tilde{\varepsilon}_{(n)}$ by choosing 
\ba
F_i^{\mu \nu} = \frac{\partial^2 F_i}{\partial \pi_\mu \partial \pi_\nu}
\ea
where $F_i$ is some scalar function of $X$ and $\pi$: such $F_i^{
\mu \nu}$ arise in the hierachical construction of \cite{Fairlie} (see section \ref{ANTI}). In this case
\ba
\frac{\partial^2 F_i}{\partial \pi_\mu \partial \pi_\nu} = 4 \frac{\partial^2 F_i}{\partial X^2} \pi^\mu \pi^\nu+  2 \frac{\partial F_i}{\partial X} g^{\mu \nu}.
\ea
Note that if there is more than one tensor $F^{\mu \nu}_{i}$ proportional to $\pi^\mu$ (or $\pi^\nu$) in Eq.~(\ref{DEFE}), then the contraction with the Levi-Civita tensor on the right hand side of this equation implies a vanishing $\CurlyE$.

Having at hand the tensors $\CurlyE$,
it is  easy to build tensors $\at$ with the required properties: namely simply replace the Levi-Civita $\varepsilon$ tensor on the right hand side of equation (\ref{DEFAten}) by the $\CurlyE$ tensor, and define, in analogy with (\ref{TENSGAL1}) a tensor $\at$ given by
\begin{equation} \label{DEFEbis}
\at_{(2n)}^{\mu_{\vphantom{()}1} \mu_{\vphantom{()}2}
\ldots \mu_{\vphantom{()}n} \nu_{\vphantom{()}1} \nu_{\vphantom{()}2}
\ldots \nu_{\vphantom{()}n}} \equiv f \times \,
\CurlyE_{\{F_i\}}^{\hphantom{\{F_i\}}\mu_{\vphantom{()}1}
\mu_{\vphantom{()}2}  \ldots
\mu_{\vphantom{()}n} \sigma_{\vphantom{()}1}\sigma_{\vphantom{()}2}\ldots
\sigma_{\vphantom{()}D-n}}
\,\CurlyE_{\{G_i\} \hphantom{\nu_{\vphantom{()}1} \nu_{\vphantom{()}2} \ldots
\nu_{\vphantom{()}n}} \sigma_{\vphantom{()}1}
\sigma_{\vphantom{()}2}\ldots \sigma_{\vphantom{()}D-n}}^{\hphantom{\{G_i\}}\nu_{\vphantom{()}1} \nu_{\vphantom{()}2} \ldots
\nu_{\vphantom{()}n}}.
\end{equation}
Here the tensors $\CurlyE$ are built with (possibly different) given sets of functions $F_{i}$ and $G_{i}$, and the function $f$ is an arbitrary scalar function of $\pi_\mu$ and $\pi$.
For example, if we choose $f=1$, one of the $\CurlyE$ tensors to be the tensor 
$\tilde{\varepsilon}_{(n)}$, and the other to be the Levi-Civita tensor $\epsilon$, then $\at$ is simply the tensor $\at_{(2n),{\rm{Gal}},2}$ of Eq.(\ref{TGAL2}), namely
\begin{equation} \label{DEFEter}
\at_{(2n)}^{\mu_{\vphantom{()}1} \mu_{\vphantom{()}2}
\ldots \mu_{\vphantom{()}n} \nu_{\vphantom{()}1} \nu_{\vphantom{()}2}
\ldots \nu_{\vphantom{()}n}} \equiv
\tilde{\varepsilon}_{(n)}^{\hphantom{(n)}\mu_{\vphantom{()}1}
\mu_{\vphantom{()}2}  \ldots
\mu_{\vphantom{()}n} \sigma_{\vphantom{()}1}\sigma_{\vphantom{()}2}\ldots
\sigma_{\vphantom{()}D-n}}
\,\varepsilon_{ \hphantom{\nu_{\vphantom{()}1} \nu_{\vphantom{()}2} \ldots
\nu_{\vphantom{()}n}} \sigma_{\vphantom{()}1}
\sigma_{\vphantom{()}2}\ldots \sigma_{\vphantom{()}D-n}}^{\nu_{\vphantom{()}1} \nu_{\vphantom{()}2} \ldots
\nu_{\vphantom{()}n}}.
\end{equation}

\label{ANTI} 
\subsection{Euler hierachies} \label{EULERHIER}
Another way to generate tensor of the form (\ref{DEFE}) is provided by the {\it Euler hierachies} of \cite{Fairlie}, which we now revisit using the notations and results of the present paper. 
We start from a set of arbitrary scalar functions $F_\ell = F_\ell(\pi^\mu)$ which depend {\it only} on first derivatives of the scalar field $\pi$, and work in flat $D$-dimensional space-time. Then, denoting $W_0 = 1$, we define the recursion relation  
\ba
W_{\ell+1}=-\hat{\mathcal{E}}F_{\ell+1}W_{\ell},
\ea
where $\hat{\mathcal{E}}$ is the Euler-Lagrange operator defined in Eq.~(\ref{ELHAT}). Hence $W_{\ell}$ is the field equation of a Lagrangian  $\mathcal{L}_{\ell}$ defined by 
\ba
\mathcal{L}_{\ell}=F_{\ell}W_{\ell-1},
\ea
and we build that way an ``Euler hierarchy" of equations of motion and Lagrangian for each $\ell$.  An interesting aspect of this hierarchy is that one can show that $W_\ell$ is given by 
\ba \label{WlWl}
W_{\ell}=\mathcal{A}_{\mu_{1}\mu_{2}\cdots\mu_{\ell}}{}^{\nu_{1}\nu_{2}\cdots\nu_{\ell}} \frac{\partial^{2} F_{1}}{\partial\pi_{\mu_{1}}\partial\pi_{\rho_{1}}}\frac{\partial^{2} F_{2}}{\partial\pi_{\mu_{2}}\partial\pi_{\rho_{2}}}\cdots  \frac{\partial^{2} F_{\ell}}{\partial\pi_{\mu_{\ell}}\partial\pi_{\rho_{\ell}}}\pi_{\rho_{1}\nu_{1}}\pi_{\rho_{2}\nu_{2}}\cdots\pi_{\rho_{\ell}\nu_{\ell}},
\ea
and
\be
\mathcal{L}_{\ell}=\mathcal{A}_{\mu_{1}\cdots\mu_{\ell-1}}{}^{\nu_{1}\cdots\nu_{\ell-1}} \frac{\partial^{2} F_{1}}{\partial\pi_{\mu_{1}}\partial\pi_{\rho_{1}}}\cdots  \frac{\partial^{2} F_{\ell-1}}{\partial\pi_{\mu_{\ell-1}}\partial\pi_{\rho_{\ell-1}}}F_{\ell}\pi_{\rho_{1}\nu_{1}}\cdots\pi_{\rho_{\ell-1}\nu_{\ell-1}} \, .
\ee
The proof proceeds by induction \cite{Fairlie}, and is also given in appendix \ref{PROOFINDUC}.  Notice that $\mathcal{L}_{\ell}$ can be obtained from (\ref{DEFEbis}) by setting $f\equiv \frac{1}{(\ell-1)!}F_{\ell}$ and contracting the tensor $\CurlyE_{\{\frac{\partial^2 F _{k}}{\partial\pi_{\mu}\partial\pi_{\nu}}\}}$ with $\varepsilon$.
The above generic hierarchy stops after {\it at most} $D$ steps (it may stop earlier depending on the properties of the functions $F_k$) and one finds a vanishing $W_{D+1}$.  As discussed in \cite{Fairlie}, and as we will see in a simple example below, the last non trivial equation of motion $W_D$ is simply given (whatever the choice made for the functions $F_k$) by the equation of motion of the maximal Galileon in $D$ dimensions, namely the one for which $N=D+1$, see (\ref{MAXX}).  

The maximal and non maximal Galileons can also be obtained by choosing $F_{k} =  \pi_{\mu}\pi^{\mu}/2$ for all ${k}=0,\ldots,\ell$. In this case, one has 
\ba
W_{\ell}&=&\mathcal{A}_{(2\ell)}^{\mu_{1}\cdots\mu_{\ell}\nu_{1}\cdots\nu_{\ell}}\pi_{\mu_{1}\nu_{1}}\cdots\pi_{\mu_{\ell}\nu_{\ell}} = \mathcal{E}_{\ell+1} \\
\mathcal{L}_{\ell}&=& \frac{1}{2} X W_{\ell-1} =\frac{1}{2}\mathcal{L}_{\ell+1}^{{\rm Gal},3}. 
\ea
More generally, when the $F_k$ are (possibly different) functions of $X$, that is $F_k= f_k(\pi^\lambda \pi_\lambda)$, one can show that 
\ba
W_{\ell}&=&\alpha_\ell \mathcal{E}_{\ell+1}-\beta_\ell \mathcal{L}_{\ell+2}^{{\rm Gal},1},
\\
{\cal L}_{\ell} &=& f_\ell(X) \left( \alpha_{\ell-1} \mathcal{E}_{\ell}-\beta_{\ell-1} \mathcal{L}_{\ell+1}^{{\rm Gal},1} \right)
\label{LEULER}
\ea
where $\alpha$ and $\beta$ are given by 
\ba
\alpha_\ell(\pi^{\lambda}\pi_{\lambda})&=& \frac{2^{\ell-1}}{\ell}\sum_{k=1}^{\ell}\left(\left(2f_{k}'+4(\pi^{\lambda}\pi_{\lambda})f_{k}''\right)\prod_{j\neq k}f_{j}'\right)
\nn
\\
\beta_\ell(\pi^{\lambda}\pi_{\lambda})&=&\frac{2^{\ell+1}}{\ell}\sum_{k=1}^{\ell}\left(f_{k}''\prod_{j\neq k}f_{j}'\right) \, .
\nn
\ea
To conclude, since both ${\cal L}_{\ell+1}^{{\rm Gal},1}$ and $\mathcal{E}_{\ell}$ are directly related to ${\cal L}_{\ell+1}^{{\rm Gal},3}$ (see equations (\ref{REL1}) and (\ref{DEFGAL3}) respectively), the Lagrangian (\ref{LEULER}) which we have constructed using the Euler hierarchies belongs to our family of general Lagrangians given in (\ref{GENL}).

\subsection{Conformal covariant Galileons in 4 dimensions}
\label{CONFGALCOV}

The conformal galileons \cite{Nicolis:2008in} provide an other simple example of theories of the kind obtained in this work (other non trivial examples in flat and curved space-times have been obtained in \cite{VanAcoleyen:2011mj}). The Lagrangians for the conformal Galileons in 4 dimensions and flat space-time have been given explicitly in  \cite{deRham:2010eu}. They read, in our notation, 
\ba
\mathcal{L}^{{\rm C.Gal}}_{4} & = & \frac{1}{20}e^{2\pi}X\left[10\left(\left(\square\pi\right)^{2}-\pi_{\mu\nu}\pi^{\mu\nu}\right)+4\left(X\square\pi-\pi^{\mu}\pi^{\nu}\pi_{\mu\nu}\right)+3X^{2}\right], \nn \\
\mathcal{L}^{{\rm C.Gal}}_{5}&=&e^{4\pi}X\left[\frac{1}{3}\left(\left(\square\pi\right)^{3}+2\pi_{\mu_{1}}^{\mu_{2}}\pi_{\mu_{2}}^{\mu_{3}}\pi_{\mu_{3}}^{\mu_{1}}-3\square\pi\pi_{\mu\nu}\pi^{\mu\nu}\right)+X\left(\left(\square\pi\right)^{2}-\pi_{\mu\nu}\pi^{\mu\nu}\right)\right.\nn \\
&&\left.+\frac{10}{7}X\left(X\square\pi-\pi^{\mu}\pi^{\nu}\pi_{\mu\nu}\right)-\frac{1}{28}X^{3}\right]. \nn
\ea
These can be rewritten in terms of $\mathcal{L}_{N}^{\text{Gal},i}$ and $\mathcal{L}_{n}^{\left(i\right)}\left\{ f\right\} $, as 
\begin{eqnarray}
\mathcal{L}^{{\rm C.Gal}}_{4} & = & -\frac{1}{2}e^{2\pi}\mathcal{L}_{4}^{\text{Gal},3}-\frac{1}{5}e^{2\pi}X\mathcal{L}_{3}^{{\rm \text{Gal},1}}+\frac{3}{20}e^{2\pi}X^{3}\nonumber, \\
 & = & -\frac{1}{2}\mathcal{L}_{2}^{\left(3\right)}\left\{ e^{2\pi}\right\} -\frac{1}{5}\mathcal{L}_{1}^{\left(1\right)}\left\{ e^{2\pi}X\right\} +\frac{3}{20}e^{2\pi}X^{3},\label{LCG4} \\
 \mathcal{L}^{{\rm C.Gal}}_{5} & = & -\frac{1}{3}e^{4\pi}\mathcal{L}_{5}^{\text{Gal},3}-e^{4\pi}X\mathcal{L}_{4}^{\text{Gal},3}-\frac{10}{7}e^{4\pi}X^{2}\mathcal{L}_{3}^{{\rm \text{Gal},1}}-\frac{1}{28}e^{4\pi}X^{4},\nonumber \\
 & = & -\frac{1}{3}\mathcal{L}_{3}^{\left(3\right)}\left\{ e^{4\pi}\right\} -\mathcal{L}_{2}^{\left(3\right)}\left\{ e^{4\pi}X\right\} -\frac{10}{7}\mathcal{L}_{1}^{\left(3\right)}\left\{ e^{4\pi}X^{2}\right\} -\frac{1}{28}e^{4\pi}X^{4}.\label{LCG5}\end{eqnarray}
As expected, one sees that those Lagrangians are indeed in the family obtained by our proof of uniqueness.  
 The covariantization of theories (\ref{LCG4}) and  (\ref{LCG5}) is straightforward.  Only the first term in (\ref{LCG4}) and the first two terms in (\ref{LCG5})
need compensating factors. The relevant covariantization formula are
\begin{eqnarray}
\mathcal{L}_{2}^{\text{cov}}\left\{ f\right\}  & = & \mathcal{L}_{2}^{\left(3\right)}\left\{ f\right\} +\frac{1}{2}R\int_{X_{0}}^{X}dY\, f\left(\pi,Y\right)Y,\label{L2_cov}
\\
\mathcal{L}_{3}^{\text{cov}}\left\{ f\right\}  & = & \mathcal{L}_{3}^{\left(3\right)}\left\{ f\right\} -3G_{\mu\nu}\pi^{\mu\nu}\int_{X_{0}}^{X}dY\, f\left(\pi,Y\right)Y.\label{L3_cov}
\end{eqnarray}
To summarize, (\ref{LCG4}) must be completed with
\begin{equation} \label{L4CGALC}
\mathcal{L}_{4}^{{\rm C.Gal,c}}=-\frac{1}{8}e^{2\pi}R\left(X^{2}-X_{0}^{2}\right),\end{equation}
and (\ref{LCG5}) must be completed with\begin{equation} \label{L5CGALC}
\mathcal{L}_{5}^{{\rm C.Gal,c}}=\frac{1}{2}e^{4\pi}\left[G_{\mu\nu}\pi^{\mu\nu}\left(X^{2}-X_{0}^{2}\right)-\frac{1}{3}R\left(X^{3}-X_{0}^{3}\right)\right],
\end{equation}
where $X_{0}$ is an arbitrary integration constant that can be taken to vanish.  
Note that the coefficients of the terms proportional to $(X^2-X_0^2)$ in (\ref{L4CGALC}) and (\ref{L5CGALC}) can be directly read from the expressions given in \cite{US,Deffayet:2009mn}. However, one should take into account that one has to integrate once $e^{4 \pi} X {\cal T}_{(4),\text{Gal},3}$ to obtain the term proportional to $(X^3-X_0^3)$ in (\ref{L5CGALC}).

\section{Conclusions}
In this work, we have obtained the most general scalar theory which has an action depending on derivatives of order up to second and has second order (and lower) field equations. Those theories were shown to have Lagrangians made by taking the product of an arbitrary function of the scalar field and its first derivatives with a special form of the Galileon Lagrangian, or any linear combinations of those Lagrangians. We have also shown how to covariantize those models, while maintaining the key property that field equations are second order. Finally, we have also discussed the relation between our construction and the Euler hierarchies of Fairlie {\it et al.}. We have shown in particular that the latter construction allows one to obtain all theories which are shift symmetric. 

Several questions are left for future work. On the formal side, it would be interesting to see how the above proof can be generalized to the case of $p$-forms and/or multifields. One could also investigate the possibility of having actions which depend on derivatives of order higher than two, and yet give rise to field equations of second order. On the phenomenological side, work is needed to see which subsets of the theories introduced here retain the several interesting aspects recalled in the introduction.  Finally, it would be interesting to study the cosmology of, as well as cosmological perturbation theory in, phenomenologically interesting models.

\section*{Acknowledgements}
C.D.~thanks S.~Deser and  G.~Esposito-Farese, as well as C.~de~Rham, J.~Khoury and A.~Tolley  for discussions. X.G.~thanks N.~Deruelle, D.~Israel, B.~Shou, J.~Troost for interesting discussions and comments. We also thank D.~Langlois for many useful discussions and for his interest at an early stage of this project.


\begin{appendix}
\section{Useful Identities}
\label{PROPA}

The definition of $\mathcal{A}_{(2n)}$ given in (\ref{DEFAten}) is
\be
\mathcal{A}_{(2n)}^{\mu_{\vphantom{()}1}\mu_{\vphantom{()}2}\ldots\mu_{\vphantom{()}n}\nu_{\vphantom{()}1}\nu_{\vphantom{()}2}\ldots\nu_{\vphantom{()}n}}\equiv\frac{1}{(D-n)!}\varepsilon_{\vphantom{\mu_{\vphantom{()}1}}}^{\mu_{\vphantom{()}1}\mu_{\vphantom{()}2}\ldots\mu_{\vphantom{()}n}\sigma_{\vphantom{()}1}\sigma_{\vphantom{()}2}\ldots\sigma_{\vphantom{()}D-n}}\,\varepsilon_{\hphantom{\nu_{\vphantom{()}1}\nu_{\vphantom{()}2}\ldots\nu_{\vphantom{()}2n}}\sigma_{\vphantom{()}1}\sigma_{\vphantom{()}2}\ldots\sigma_{\vphantom{()}D-n}}^{\nu_{\vphantom{()}1}\nu_{\vphantom{()}2}\ldots\nu_{\vphantom{()}n}},
\label{OPERA}
\ee
where (see (\ref{DEFLC}))
\begin{eqnarray*}
\varepsilon^{\mu_{\vphantom{()}1}\mu_{\vphantom{()}2}\ldots\mu_{\vphantom{()}D}} & = & -\frac{1}{\sqrt{-g}}\delta_{1}^{[\mu_{\vphantom{()}1}}\delta_{2}^{\mu_{\vphantom{()}2}}\ldots\delta_{D}^{\mu_{\vphantom{()}D}]}\equiv-\frac{1}{\sqrt{-g}}\delta_{12\cdots D}^{\mu_{1}\mu_{2}\cdots\mu_{D}} \, ,
\\
\varepsilon_{\mu_{1}\mu_{2}\cdots\mu_{D}} & = &
 {\sqrt{-g}}\delta^{1}_{[\mu_{\vphantom{()}1}}\delta^{2}_{\mu_{\vphantom{()}2}}\ldots\delta^{D}_{\mu_{\vphantom{()}D}]}
 \equiv  \sqrt{-g}\delta_{\mu_{1}\mu_{2}\cdots\mu_{D}}^{12\cdots D},
\end{eqnarray*}
so that (\ref{OPERA}) can be rewritten as
 \begin{eqnarray}
\mathcal{A}_{\left(2n\right)\phantom{\mu_{1}\mu_{2}\cdots\mu_{n}}\nu_{1}\nu_{2}\cdots\nu_{n}}^{\phantom{\left(2n\right)}\mu_{1}\mu_{2}\cdots\mu_{n}} 
\nonumber
& = & 
-\frac{1}{(D-n)!}\delta_{\nu_{\vphantom{()}1}\nu_{\vphantom{()}2}\ldots\nu_{\vphantom{()}n}\sigma_{\vphantom{()}1}\sigma_{\vphantom{()}2}\ldots\sigma_{\vphantom{()}D-n}}^{\mu_{\vphantom{()}1}\mu_{\vphantom{()}2}\ldots\mu_{\vphantom{()}n}\sigma_{\vphantom{()}1}\sigma_{\vphantom{()}2}\ldots\sigma_{\vphantom{()}D-n}}
  \nonumber \\
 & = & -
 \delta_{\nu_{1}\cdots\nu_{n}}^{\mu_{1}\cdots\mu_{n}}
 \label{USEFUL1}
 \end{eqnarray}
The identity between the three Galileon Lagrangians (\ref{LGAL1}), (\ref{LGAL2}) and (\ref{LGAL3}) given in (\ref{3Lid}) then follows by using
\ba
\delta_{\nu_{1}\nu_{2}\cdots\nu_{n+1}}^{\mu_{1}\mu_{2}\cdots\mu_{n+1}}&=&\sum_{i=1}^{n+1}\left(-1\right)^{i-1}\delta_{\nu_{i}}^{\mu_{1}}\delta_{\nu_{1}\nu_{2}\cdots\nu_{i-1}\nu_{i+1}\dots\nu_{n+1}}^{\mu_{2}\mu_{3}\cdots\mu_{n+1}}
\nonumber
\\
&=& \delta_{\nu_{1}}^{\mu_{1}}\delta_{\nu_{2}\cdots\dots\nu_{n+1}}^{\mu_{2}\cdots\mu_{n+1}}
+ \sum_{i=2}^{n+1}\left(-1\right)^{i-1}\delta_{\nu_{i}}^{\mu_{1}}\delta_{{\nu_{1}}\nu_{2}\cdots\nu_{i-1}\nu_{i+1}\dots\nu_{n+1}}^{\mu_{2}\mu_{3}\cdots\mu_{i-1}{\mu_{i}}\mu_{i+1}\cdots\mu_{n+1}} .
\label{USEFUL2}
\ea

\section{A formal proof of the first lemma a of section \ref{LEM1}}
\label{APPA}
Here we provide a formal proof of the lemma of section \ref{LEM1}. Since we also use this lemma in section \ref{COVAR} we discuss the proof in curved space-time with the understanding that we replace everywhere partial derivatives $\partial$ by covariant derivatives $\nabla$ (i.e. $\pi_{\alpha}\equiv \nabla_{\alpha}\pi$, $\pi_{\alpha\beta}\equiv \nabla_{\alpha}\nabla_{\beta}\pi$ and so on). We consider
${\cal T}_{(2)}$, a twice contravariant tensor depending on $\pi$ and its first derivatives $\pi_\mu$. Varying the Lagrangian ${\cal L}= {\cal T}^{\mu \nu}_{(2)} \pi_{\mu \nu}$, and writing only
the potentially dangerous  terms (i.e. those terms which can lead to third and higher derivatives, which, following the notations of \cite{Deffayet:2009mn}, we denote by using the symbol $\sim$) in the variations we find (using suitable integrations by part)
\ba
\delta\mathcal{L}&=& \frac{\partial {\cal T}_{(2)}^{\mu\nu}}{\partial \pi_{\rho}}\delta\pi_{\rho}\pi_{\mu\nu}+{\cal T}_{(2)}^{\mu\nu}\delta\pi_{\mu\nu} \nn \\
&\sim&  -\frac{\partial {\cal T}_{(2)}^{\mu\nu}}{\partial \pi_{\rho}}\pi_{\rho\mu\nu}\delta\pi+\left(\nabla_{\nu}\nabla_{\mu}{\cal T}_{(2)}^{\mu\nu}\right)\delta\pi\nn\\
&\sim&  -\frac{\partial {\cal T}_{(2)}^{\mu\nu}}{\partial \pi_{\rho}}\pi_{\rho\mu\nu}\delta\pi+\left(\partial_{\nu}\partial_{\mu}{\cal T}_{(2)}^{\mu\nu}\right)\delta\pi\nn\\
&\sim&  -\frac{\partial {\cal T}_{(2)}^{\mu\nu}}{\partial \pi_{\rho}}\pi_{\rho\mu\nu}\delta\pi+\partial_{\nu}\left(\frac{\partial {\cal T}_{(2)}^{\mu\nu}}{\partial\pi_{\rho}}\partial_{\mu}\pi_{\rho}\right)\delta\pi\nn\\
&\sim&  -\frac{\partial {\cal T}_{(2)}^{\mu\nu}}{\partial \pi_{\rho}}\pi_{\rho\mu\nu}\delta\pi+\frac{\partial {\cal T}_{(2)}^{\mu\nu}}{\partial\pi_{\rho}}\partial_{\nu}\partial_{\mu}\pi_{\rho}\delta\pi\nn\\
&\sim&  -\frac{\partial {\cal T}_{(2)}^{\mu\nu}}{\partial \pi_{\rho}}\left(\pi_{\rho\mu\nu}- \pi_{\nu\mu\rho}\right)\delta\pi\nn
\ea
Since the commutation of the covariant derivatives only involves a contraction of the Riemann tensor with a first derivative of the field, the dangerous terms exactly cancel each other and this concludes the proof.

\section{Euler Hierachy}
\label{PROOFINDUC}
In this appendix, following \cite{Fairlie}, we show how to obtain (\ref{WlWl}).
The proof proceeds by induction. First notice that
$$W_{1}=-\hat{\mathcal{E}}F_{1}W_{0}=\partial_{\mu}\frac{\partial F_{1}}{\partial\pi_{\mu}}=\frac{\partial^{2}F_{1}}{\partial \pi_{\mu}\partial\pi_{\nu}}\pi_{\mu\nu}\, ,$$
so the proposition is true for $n=1$. Suppose now that it is true for arbitrary fixed $n$. In order to show that it is still valid for $n+1$ we will need the following identity
\begin{align*}
\hat{\mathcal{E}}F(\pi_{\mu})W(\pi_{\mu},\pi_{\mu\nu}) = \left(\hat{\mathcal{E}}W\right)F &-2\left[\frac{\partial W}{\partial\pi_{\mu}}-\frac{1}{2}\partial_{\nu}\left(\frac{\partial W}{\partial\pi_{\mu\nu}}+\frac{\partial W}{\partial\pi_{\nu\mu}}\right)\right]\pi_{\mu\rho}\frac{\partial F}{\partial \pi_{\rho}} \\
 &- \left(W\frac{\partial^{2}F}{\partial \pi_{\mu}\partial \pi_{\nu}}\pi_{\mu\nu}-\frac{\partial W}{\partial\pi_{\mu\nu}}\frac{\partial^{2}F}{\partial \pi_{\rho}\partial \pi_{\sigma}}\pi_{\mu\rho}\pi_{\nu\sigma}\right)\, .
\end{align*}
We can now use this with $W=W_{n}$ (which depends only on first and second derivatives of $\pi$) and $F=F_{n+1}$. But since $W_{n}$ is a divergence (it is the equation of motion of some lagrangian which does not depend explicitly on the field itself), its equation of motion vanishes $\hat{\mathcal{E}}W_{n}=0$ (alternatively we could say that the operator $\hat{\mathcal{E}}$ is nilpotent). Also because of the antisymmetry of the Levi-Civita tensor, we can show that
\begin{align*}
\frac{\partial W_{n}}{\partial\pi_{\mu}}-\frac{1}{2}\partial_{\nu}\left(\frac{\partial W_{n}}{\partial\pi_{\mu\nu}}+\frac{\partial W_{n}}{\partial\pi_{\nu\mu}}\right)=\frac{\partial W_{n}}{\partial\pi_{\mu}} 
&-\frac{1}{2}\underbrace{\left(\frac{\partial^{2}W_{n}}{\partial \pi_{\mu\nu}\partial\pi_{\rho}}+\frac{\partial^{2}W_{n}}{\partial \pi_{\nu\mu}\partial\pi_{\rho}}\right)\pi_{\nu\rho}}_{=2\frac{\partial W_{n}}{\partial\pi_{\mu}}}\\
&+\frac{1}{2}\underbrace{\left(\frac{\partial^{2}W_{n}}{\partial \pi_{\mu\nu}\partial\pi_{\rho\sigma}} +\frac{\partial^{2}W_{n}}{\partial \pi_{\nu\mu}\partial\pi_{\rho\sigma}}\right)\pi_{\nu\rho\sigma}}_{=0}\, .
\end{align*}
So in the end this term vanishes too and we have
\begin{align*}
-\hat{\mathcal{E}}F_{n+1}W_{n}& = W_{n}\frac{\partial^{2}F_{n+1}}{\partial \pi_{\mu}\partial \pi_{\nu}}\pi_{\mu\nu}-\frac{\partial W_{n}}{\partial\pi_{\mu\nu}}\frac{\partial^{2}F_{n+1}}{\partial \pi_{\rho}\partial \pi_{\sigma}}\pi_{\mu\rho}\pi_{\nu\sigma} \\
&=\mathcal{A}_{\mu_{1}\cdots\mu_{n}}{}^{\nu_{1}\cdots\nu_{n}} \frac{\partial^{2} F_{1}}{\partial\pi_{\mu_{1}}\partial\pi_{\rho_{1}}} \cdots  \frac{\partial^{2} F_{n}}{\partial\pi_{\mu_{n}}\partial\pi_{\rho_{n}}}\frac{\partial^{2} F_{n+1}}{\partial\pi_{\mu}\partial\pi_{\nu}}\pi_{\rho_{1}\nu_{1}}\cdots\pi_{\rho_{n}\nu_{n}}\pi_{\mu\nu}\\
&\ \ \ \ -\sum_{p=1}^{n}\mathcal{A}_{\mu_{1}\cdots\mu_{n}}{}^{\nu_{1}\cdots\nu_{p-1}\nu\nu_{p}\cdots\nu_{n}} \frac{\partial^{2} F_{1}}{\partial\pi_{\mu_{1}}\partial\pi_{\rho_{1}}}\cdots \frac{\partial^{2} F_{p}}{\partial\pi_{\mu_{p}}\partial\pi_{\mu}} \cdots
  \frac{\partial^{2} F_{n}}{\partial\pi_{\mu_{n}}\partial\pi_{\rho_{n}}} \frac{\partial^{2}F_{n+1}}{\partial \pi_{\rho}\partial \pi_{\sigma}} \\
  &\ \ \ \ \ \ \ \ \ \ \ \ \ \ \ \ \ \ \ \ \ \ \ \ \ \ \ \ \ \ \ \  \times \pi_{\rho_{1}\nu_{1}}\cdots \pi_{\rho_{p-1}\nu_{p-1}}\pi_{\mu\rho}\pi_{\nu\sigma} \pi_{\rho_{p+1}\nu_{p+1}}\cdots\pi_{\rho_{n}\nu_{n}} \\
  & =  \left(\mathcal{A}_{\mu_{1}\cdots\mu_{n}}{}^{\nu_{1}\cdots\nu_{n}}\delta_{\mu_{n+1}}{}^{\nu_{n+1}}-\sum_{p=1}^{n}\mathcal{A}_{\mu_{1}\cdots\mu_{n}}{}^{\nu_{1}\cdots\nu_{p-1}\nu_{n+1}\nu_{p+1}\cdots\nu_{n}}\delta_{\mu_{n+1}}{}^{\nu_{p}} \right) \\
  &\ \ \ \ \ \ \ \ \ \ \ \ \ \ \ \ \ \ \ \ \ \ \ \ \ \ \ \ \ \ \ \  \times \frac{\partial^{2} F_{1}}{\partial\pi_{\mu_{1}}\partial\pi_{\rho_{1}}}\cdots \frac{\partial^{2} F_{n+1}}{\partial\pi_{\mu_{n+1}}\partial\pi_{\rho_{n+1}}}\pi_{\rho_{1}\nu_{1}}\cdots\pi_{\rho_{n+1}\nu_{n+1}}\, .
\end{align*}
But we also have (with \ref{USEFUL2})
$$\mathcal{A}_{\mu_{1}\cdots\mu_{n}}{}^{\nu_{1}\cdots\nu_{n}}\delta_{\mu_{n+1}}{}^{\nu_{n+1}}-\sum_{p=1}^{n}\mathcal{A}_{\mu_{1}\cdots\mu_{n}}{}^{\nu_{1}\cdots\nu_{p-1}\nu_{n+1}\nu_{p+1}\cdots\nu_{n}}\delta_{\mu_{n+1}}{}^{\nu_{p}} = \mathcal{A}_{\mu_{1}\cdots\mu_{n+1}}{}^{\nu_{1}\cdots\nu_{n+1}}\, ,$$
which proves our result for $n+1$. Notice that in this proof, we often used the fact that partial derivatives commute over flat space-time.

\end{appendix}

\end{document}